\documentclass[%
 reprint,
 amsmath,amssymb,
 aps,
]{revtex4-2}

\newcommand{\RNum}[1]{\uppercase\expandafter{\romannumeral #1\relax}}
\usepackage{siunitx}
\usepackage{graphicx}
\usepackage{dcolumn}
\usepackage{bm}
\usepackage{comment}

\usepackage{textcomp}
\usepackage{siunitx}
\usepackage{xcolor}
\usepackage{array}
\usepackage{multirow}

\renewcommand{\figurename}{\textbf{Fig.}}

\makeatletter
\renewcommand*{\fnum@figure}{{\normalfont\bfseries \figurename~\thefigure}}
\makeatother

\begin{document}

\preprint{APS/123-QED}

\title{Electronic-photonic circuit crossings}

\author{Babak Vosoughi Lahijani$^{1,2}$}\email{bala@dtu.dk}
\author{Marcus Albrechtsen$^1$}
\author{Rasmus E. Christiansen$^{2,3}$}
\author{Christian A. Rosiek$^1$}
\author{Konstantinos Tsoukalas$^1$}
\author{Mathias T. Sutherland$^1$}
\author{S{\o}ren Stobbe$^{1,2}$}

\affiliation{%
$^1$DTU Electro, Department of Electrical and Photonics Engineering, Technical University of Denmark, Ørsteds Plads 343, DK-2800 Kgs.\ Lyngby, Denmark
}%

\affiliation{%
$^2$NanoPhoton - Center for Nanophotonics, Technical University of Denmark, Ørsteds Plads 345A, DK-2800 Kgs.\ Lyngby, Denmark
}%

\affiliation{%
$^3$Department of Mechanical Engineering, Solid Mechanics, Technical University of Denmark, Nils Koppels Allé, B. 404, DK-2800 Kgs.\ Lyngby, Denmark
}%

\maketitle

\textbf{Electrical control of light in integrated photonics is central to a wide range of research and applications. It is conventionally achieved with thermo-optic tuning, but this suffers from high energy consumption and crosstalk. Nanoelectromechanical photonics could resolve these issues, but integrating this technology with conventional multilayer metal architectures is challenging, and conventional approaches do not allow crossings of electrical wires and photonic waveguides. Here, we use topology optimization to devise a single-layer electronic-photonic circuit crossing with up to \SI{99.8}{\percent} optical transmission across a \SI{20}{nm} electrical isolation trench. We focus our experiments on \SI{100}{nm} trenches and measure an average transmission of \SI{92.9}{\percent} over a \SI{100}{nm} bandwidth, in excellent agreement with theory. We use these concepts to demonstrate a monolithic silicon nanoelectromechanical add-drop switch in which the flow of photons, electrons, and mechanical motions are fully integrated within the same layer. Our work addresses an important challenge in incorporating opto-electro-mechanical topologies into photonic integrated circuits and may lead to new functionalities in nano-opto-electro-mechanical systems, optomechanics, and integrated quantum photonics.}

Information technologies rely on controlling two fundamentally different particles: electrons are strongly interacting massive charged fermions while photons are massless and charge-neutral bosons that do not interact. These differences underpin the development of optical communication systems based on low losses and high coherence as well as electronic computing relying on strong interactions. For many reasons, however, integrating electronics and photonics is becoming an increasingly important research frontier: First, a sizeable fraction of the energy consumption in modern computing is spent on routing information through lossy electrical leads, which may be significantly reduced by chip-level optical interconnects \cite{Miller:09,Ayata:2017}. Second, the developments in hybrid quantum technologies call for high-fidelity interfaces between stationary and flying qubits \cite{Elshaari:20,Lodahl:15}, which require unprecedented control and integration of electronics and photonics. Third, the integration of photonics and electronics is at the heart of a wide range of new technologies that rely on advanced nanofabrication for new applications beyond electronics \cite{Waldrop:2016}, ranging from smart sensors \cite{Rogers:21} over ultrahigh-bandwidth optical networks \cite{Levy:10} to chip-scale particle accelerators \cite{Sapra:20}, and nano-opto-electro-mechanical systems (NOEMS) \cite{Midolo2018,Haffner:2019,Ming_Wu:2022,Pierre:2021,Quack:2023,Kim:2023}.

Simultaneously controlling the flow of electrons and photons inside the same microchip is challenging because of the conflicting requirements for electronics and photonics: Electronic wires rely on conduction by free electrons but they in turn lead to significant optical losses. Therefore, both research and technology has so far resorted to various multi-layer chip architectures aiming to avoid crossings between wires and waveguides. Existing solutions include optical layers embedded in transparent insulators with metallic surface electrode layers or numerous patterned conducting layers buried below the optical layers and contacted by deep vias in order to route the electrical signals without impeding the optical performance. Common to existing solutions is the need to add more layers or processes as the circuits grow. In other words, the circuit complexity scales superlinearly with the number of (avoided) electronic-photonic crossings. While these solutions can be employed to integrate photonic circuits with CMOS chips \cite{Miller:2000,Atabaki2018,Stojanovic:18}, 
they are now a limitation for a wide range of research including routing of single photons from quantum dots \cite{Papon:19}, chip-scale LiDAR solutions \cite{Trocha:2018,Rogers:21}, photonic interconnects for 5G networks \cite{Sabella:2020}, compact sensors \cite{Yang:21}, electro-optic and thermo-optic devices \cite{Arrazola:21,Souza:18,Zhang:2021}, and nanoelectromechanical photonic switches \cite{Seok:16}. The solution proposed in this work overcomes the system-level complexity in routing electrons and photons inside the chip by introducing a carefully engineered and inversely designed monolithic device. It allows crossing electronic connections and photonic waveguides using the same device layer without introducing significant electrical resistance or optical losses. We focus our effort on silicon membranes fabricated on the silicon on insulator (SOI) platform but the concepts can be readily applied to all other semiconductor platforms.
\begin{figure*}
\centering\includegraphics[width=0.9\linewidth]{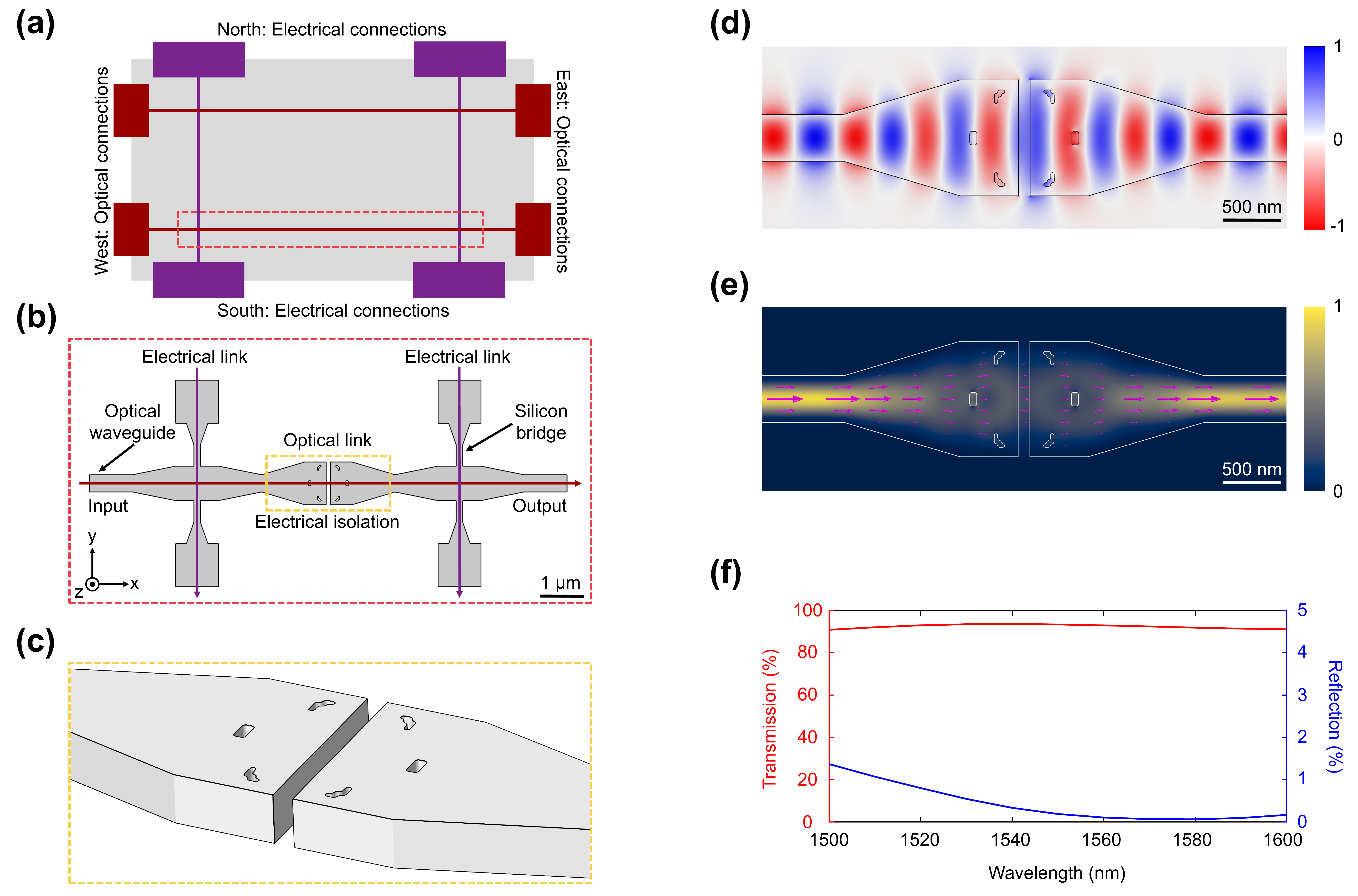}
\caption{\label{fig:Fig1}\textbf{Electronic-photonic integration with electronic-photonic circuit crossings.} \textbf{(a)} Circuit topology for generic single-layer electronic-photonic integration where electrical wires (purple) must cross photonic waveguides (red). \textbf{(b)} Schematic representation of our electronic-photonic circuit crossing. The photonic waveguide is suspended using silicon bridges, which connect the waveguide to the surrounding bulk, thus bridging current (purple arrows) across the electrical isolation trenches while keeping the optical in- and output electrically isolated. \textbf{(c)} Zoomed view of the topology-optimized EPCC. \textbf{(d)} Normalized transverse electric (TE) field ($E_\text{y}$) distribution across the optimized EPCC at \SI{1550}{nm}. \textbf{(e)} Magnitude and direction of the time-averaged power flow in the middle plane of the structure crossing a \SI{100}{nm} isolation gap at \SI{1550}{nm}. \textbf{(f)} Calculated transmission (red) and reflection (blue) of the optimized EPCC over \SI{100}{nm} wavelength span centered at \SI{1550}{nm}. Numerical calculations were performed for a device with a thickness of \SI{240}{nm} determined by the measured thickness of the device layer of the SOI wafer after fabrication. }
\end{figure*}

Figure~\ref{fig:Fig1}a illustrates a generic circuit topology for routing electrons and photons where photons are guided through the optical links, i.e., photonic waveguides. The optical ports are oriented east-west (EW) while the electrical contacts are oriented north-south (NS). Except for the most primitive integrated electronic-photonic ciruits, it is impossible to build such architectures without introducing either multilayer technologies or resorting to circuit topologies that involve crossings of photonic waveguides and electronic wires. The immediately apparent challenge of designing an electronic-photonic circuit crossing (EPCC) is to achieve both high optical transmission and low reflection in the EW direction as well as low electrical resistance in the NS direction. Furthermore, an important additional requirement for generic multiport devices is that the electrical resistance must be very high in the EW direction to avoid short circuits between the electrical leads. A simple solution would be to introduce an isolation trench through the waveguide to provide electrical isolation but this is clearly a sub-optimal solution as it significantly reduces the optical transmission.
For example, introducing a gap of \SI{100}{nm} in a suspended waveguide results in a transmission of less than \SI{32}{\percent} and a reflection above \SI{22}{\percent} over a \SI{100}{nm} bandwidth centered at \SI{1550}{nm} (fig.~S1). Alternative approaches relying on parameter-optimized tapering (fig.~S2) is also resulting in so poor performance that is detrimental to devices with even a minimum of system-level complexity.

\begin{figure*}
\centering\includegraphics[width=0.9\linewidth]{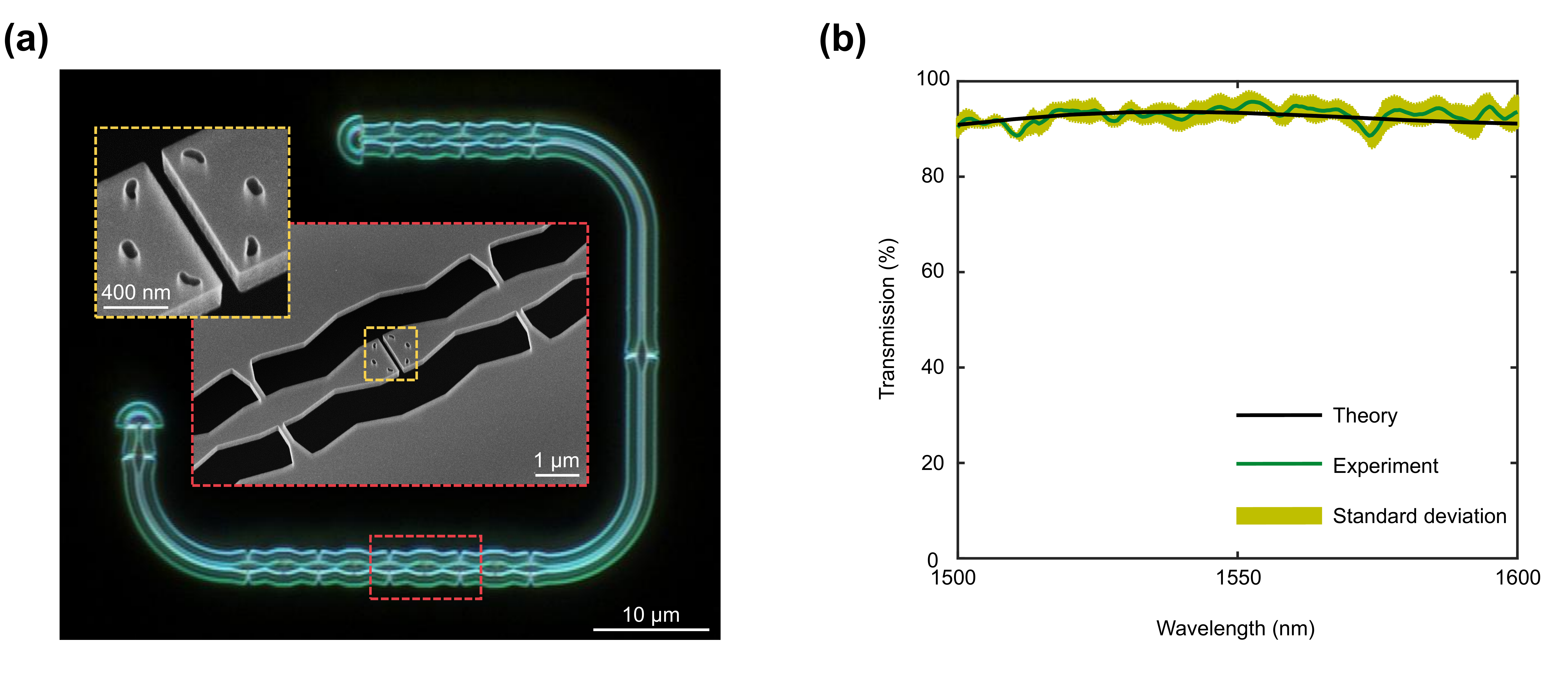}
\caption{\label{fig:Fig2}\textbf{Experimental realization of the electronic-photonic circuit crossing.} \textbf{(a)} Dark-field microscope image of a photonic circuit including 6 EPCCs fabricated on an SOI wafer. The red inset at the center shows a scanning electron micrograph (\SI{30}{\degree} tilted) of a single EPCC. The yellow inset shows the center of the EPCC. \textbf{(b)} Measured optical transmission (green) of the optimized EPCC obtained by a cut-back method showing excellent agreement with theory (black). The green shaded area represents the standard deviation from the mean of six nominally identical sets of devices.
}
\end{figure*}

An EPCC solution must ensure proper mode matching across the isolation trench. Due to the coherent nature of the photon scattering and interference, designing such a subwavelength mode-matching EPCC is a complicated and non-intuitive problem. We therefore employ inverse design by topology optimization \cite{BOOK_TOPOPT_BENDSOE,CHRISTIANSEN_SIGMUND_COMSOL_2020} to develop a compact and scalable device that transmits photons efficiently across the electrical isolation trench. The trench must be wide enough to avoid electric breakdown due to Townsend avalanches and field emission. For our experiments, we consider \SI{100}{nm} trenches that provide electrical isolation for a voltage of at least \SI{25}{V} \cite{Peschot:2014}, suitable for high-voltage applications including NOEMS.  
As we are interested in maximizing the transmission across a \SI{100}{nm} isolation trench, while at the same time minimizing reflections, we seek to maximize the figure of merit, $\Phi$ = $T$/($R$+$\alpha$), where $T$ denotes transmission across the isolation trench and $R$ the reflection, over a spatial region of $\SI{1.5}{\micro\meter}\times\SI{1.}{\micro\meter}$ that corresponds to the length and width of the design domain. The parameter $\alpha=1$ is added to the denominator to avoid numerical issues as $R\rightarrow0$ caused by the singularity at $R=0$ as well as to balance the priority to $T$ vs $R$ such that the optimization algorithm will not sacrifice $T$ when $R<<1$. The EPCC is optimized at the wavelengths corresponding to the beginning and the end of the targeted bandwidth, i.e., \SI{1500}{nm} and \SI{1600}{nm}, in order to ensure the efficient performance across a \SI{100}{nm} bandwidth centered at \SI{1550}{nm}.  

Figure~\ref{fig:Fig1}b illustrates our solution to the EPCC problem. The photonic waveguide is suspended by silicon bridges designed to provide mechanical support without disturbing the propagating optical mode, while at the same time serving as electrical links in the NS-direction. The central part is shown in Fig.~\ref{fig:Fig1}c and is generated by topology optimization, which results in excellent transmission. This is evident from the lack of resonances in the electric field (Fig.~\ref{fig:Fig1}d) and time-averaged power flow (Fig.~\ref{fig:Fig1}e) of the transverse-electric mode. The calculated transmission and reflection spectra are shown in Fig.~\ref{fig:Fig1}f and feature a transmission of $T>\SI{90.9}{\%}$ and a reflection of $R<\SI{1.4}{\%}$ over a \SI{100}{nm} bandwidth as well as a peak transmission of $T=\SI{93.6}{\%}$, which is a dramatic improvement compared to the solutions before topology optimization (figs.~S1 and S2).

\begin{figure*}
\centering\includegraphics[width=0.9\linewidth]{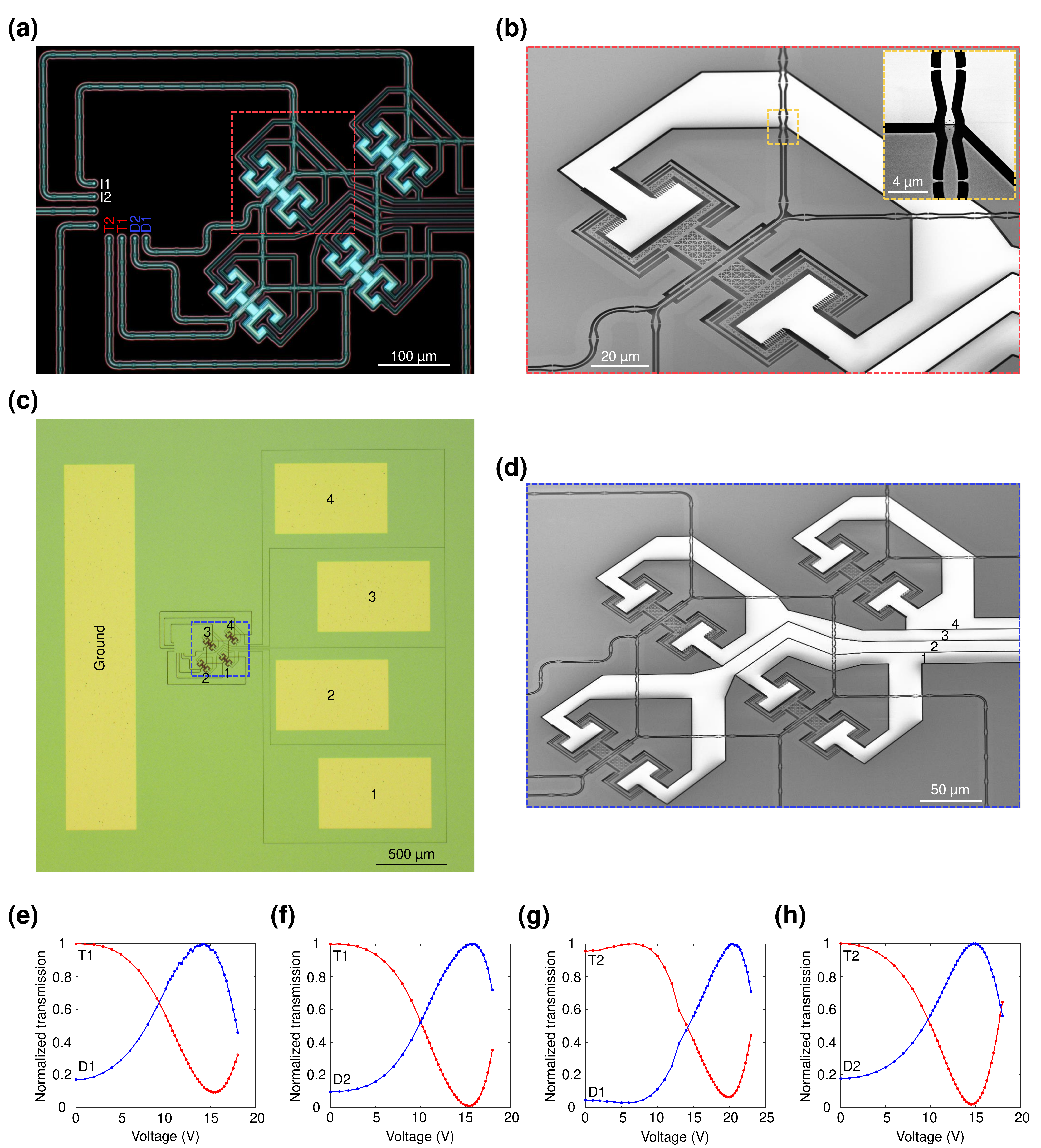}
\caption{\label{fig:Fig3} \textbf{Demonstration of a single-layer $2\times2$ add-drop switching network combining electronic, photonic, and mechanical degrees of freedom.} \textbf{(a)} Dark-field microscope image of the fabricated $2\times2$ add-drop switch made of four electrostatically actuated directional couplers. \textbf{(b)} \SI{45}{\degree}-tilted scanning electron micrograph of one switching element acquired at a low accelerating voltage of \SI{5}{kV} in order to illustrate the charging effect, showing that excellent electrical isolation is obtained between the actuator electrodes via multiple EPCCs and isolation trenches. The inset in (b) shows a zoomed top view scanning electron micrograph of one of the EPCCs employed to route the actuation signal to the electrostatic actuators. \textbf{(c)} Electrical chip layout of the switch network. Electrical pads are placed surrounding the photonic network and the voltages are routed towards four nanoelectromechanical directional couplers through multiple EPCCs. \textbf{(d)} \SI{45}{\degree}-tilted scanning electron micrograph of the four switching elements taken at a low accelerating voltage of \SI{5}{kV}. The charging effect in the SEM clearly shows how the EPCCs route the voltages from the contact pads to their corresponding nanoelectromechanical directional coupler along mutually isolated electrical domains. \textbf{(e-h)} Measured normalized transmission to through ports (red) and drop ports (blue) at \SI{1550}{nm} over actuation voltages. The incident light is coupled to $I_\text{1}$ in (e) and (f) and to $I_\text{2}$ in (g) and (h). The driving voltage is applied to the actuators at the top-left (e), the bottom-left (f), the top-right (g), and the bottom-right (h).} 
\end{figure*}

In order to experimentally verify these findings, we fabricate a set of photonic circuits with 0 to 6 EPCCs in series using electron-beam lithography and cryogenic deep reactive ion etching applied to SOI wafers. This enables direct measurement of the insertion loss of a single EPCC using the cut-back method, which separates the coupling losses from propagation losses. Figure~\ref{fig:Fig2}a shows a dark-field microscope image of such a circuit comprised of 6 EPCCs embedded in a single-mode waveguide, 20 silicon bridges, and two cross-polarized input/output couplers. The position and number of silicon bridges are chosen to provide mechanical stability for maintaining the photonic circuit suspended.
Six different sets of each photonic circuit are fabricated on the same chip, which allows us to do ensemble measurements and carefully estimate the standard deviation. Figure~\ref{fig:Fig2}b demonstrates an excellent agreement between the experimental transmission, measured over a \SI{100}{nm} bandwidth with an average transmission of \SI{92.9}{\percent}, and the numerical calculations.

\begin{figure}
\includegraphics[width=\linewidth]{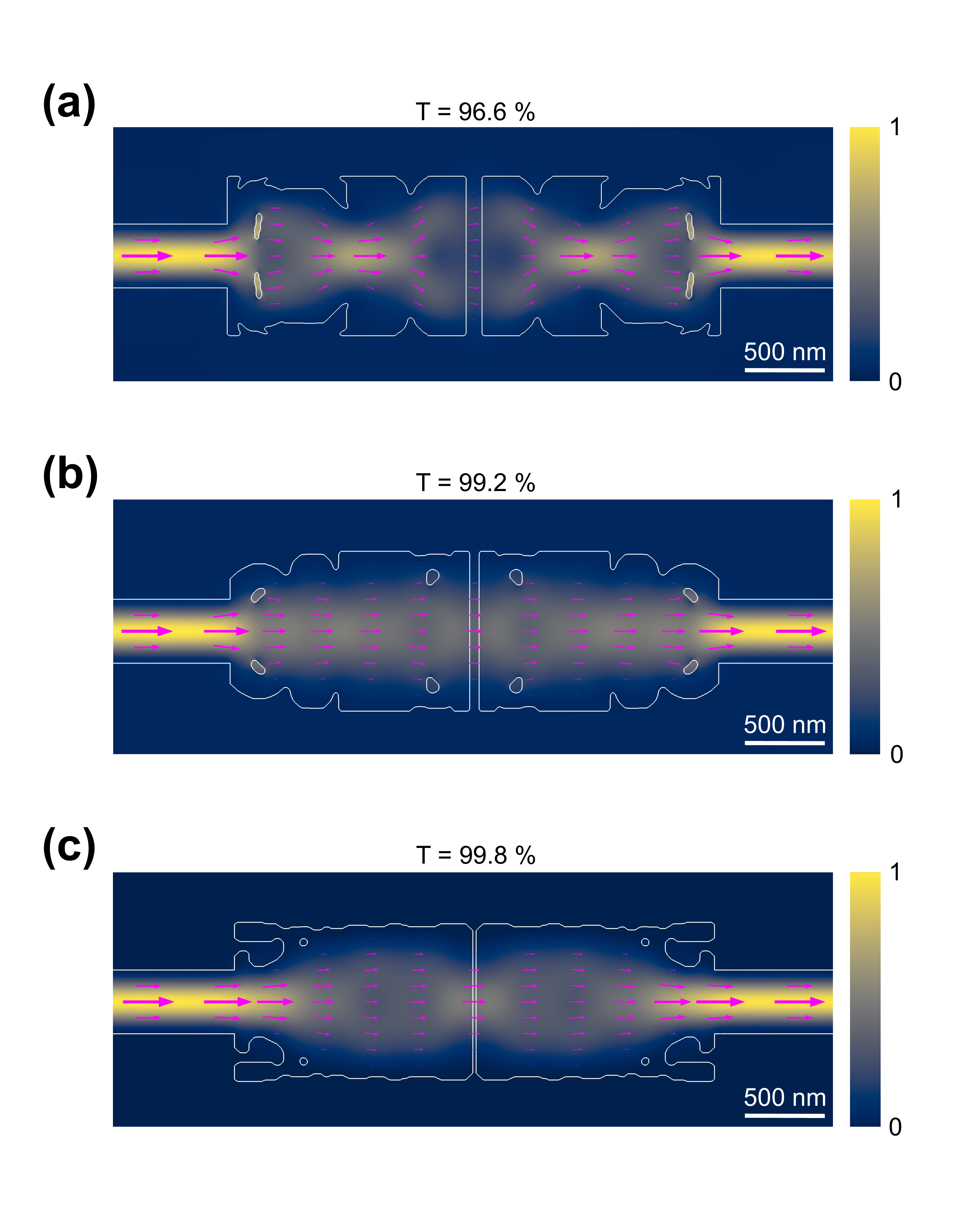}
\caption{\label{fig:Fig4} \textbf{Topology-optimized EPCCs with near-unity transmission.} The Magnitude and direction of the time-averaged power flow in the middle plane of the structure is shown for designs obtained with single-wavelength topology optimization with gaps of (\textbf{a}) \SI{100}{nm}, (\textbf{b}) \SI{60}{nm}, and (\textbf{c}) \SI{20}{nm}. The power transmission at \SI{1550}{nm} for each design is indicated in the figure and reaches \SI{99.8}{\%} for a gap of \SI{20}{nm}.} 
\end{figure}

To illustrate the application of the EPCC, we consider a new type of monolithic electro-mechanical photonic add-drop network that integrates EPCCs with nanoelectromechanical photonic switches in a single lithographic step. Figure~\ref{fig:Fig3}a shows a dark-field microscope image of the $2\times2$ switch network. The network is built on four reconfigurable directional couplers controlled by electrostatic comb-drive actuators with folded springs \cite{Legtenberg:1996,Ikeda:2013} (Fig.~\ref{fig:Fig3}b), which can induce nanoscale displacements \cite{Tsoukalas:20} to the mechanically compliant waveguides and thereby change the configuration of the switching network. Figure~\ref{fig:Fig3}b shows a scanning electron micrograph of one of the switching elements and illustrates how isolation trenches along with the EPCCs are introduced to the device layer of the SOI wafer to route the electrical signals to the comb-drive actuators. The electrical chip layout of the network is shown in Fig.~\ref{fig:Fig3}c and includes one ground and four drive contacts for actuating the switching elements. The drive contacts are electrically connected to the switching elements through the silicon device layer and isolated from each other by the isolation trenches, the EPCCs, and the supporting buried-oxide layer, as shown in Fig.~\ref{fig:Fig3}d. The switch network is
characterized by sequentially applying voltages to the drive contacts while monitoring the optical transmission from the input to the output ports. The light coupled to the input $I_\text{1}$ ($I_\text{2}$) can be transmitted to the corresponding through port $T_\text{1}$ ($T_\text{2}$) or directed to any of the drop ports $D_\text{1}$ and $D_\text{2}$ depending on the driving voltage of the activated comb-drive actuator. The through- and drop-port transmission at \SI{1550}{nm} is plotted in Fig.~\ref{fig:Fig3}e-h and show a clear switching behavior for all voltages. This demonstrates one of the central building blocks for programmable \cite{Bogaerts:20,Carolan:2015,Kim:2023} or quantum photonic circuits \cite{Arrazola:21} where the full benefit of the EPCC is ultimately harvested in larger networks: Regardless of topology and complexity, advanced electronic-photonic circuits can be fabricated  in the same device layer and in a single lithography step.

For larger networks, the optical transmission becomes a limiting factor and we therefore design additional EPCCs while relaxing the bandwidth requirement and allowing for smaller but still feasible gaps \cite{Albrechtsen:21}. Figure~\ref{fig:Fig4} shows EPCCs for (a) \SI{100}{nm}, (b) \SI{60}{nm}, and (c) \SI{20}{nm} isolation gaps topology-optimized for maximum transmission at \SI{1550}{nm}. The transmission increases with decreasing gap size and for a
gap of \SI{20}{nm}, the peak power transmission reaches a remarkably high value of $T=\SI{99.8}{\%}$. Furthermore, it remains above \SI{99}{\%} across a \SI{35}{\nm} bandwidth (fig.~S6). Because the overall transmission of circuits with $N$ crossings scales as $T^N$, a power transmission of $T=\SI{99.8}{\%}$ implies that hundreds of EPCCs in series would result in insertion losses below \SI{3}{dB}. This shows that highly complex hybrids of nanoelectronics, nanophotonics, and nanomechanics can be integrated with negligible losses. Our results indicate that near-unity transmission is attainable for smaller gaps.

The EPCC opens perspectives for a wide range of new research and technology. For example, isolation trenches are not only efficient electrical isolators, they are also ideal for thermal isolation in nanophotonic devices \cite{Souza:18} or mechanical isolation for self-assembly of nanophotonic circuits \cite{Babar:2023} and quantum optomechanics \cite{Rossi:2018}, and we envision our technology enabling, e.g., phonon-photon crossings. In addition, our concepts apply directly to other semiconductor platforms such as gallium arsenide or indium phosphide, where they enable independent electrical control of quantum emitters in integrated photonic circuits \cite{Papon:19}, thus solving a major scaling problem in integrated quantum photonics. Previous research on this subject has used shallow etching of individual doped layers for individual electrical tuning of quantum dots \cite{Papon:2023}, but this approach would not work for homogeneously doped semiconductor platforms and does not address the more general question of how to provide electrical, thermal, and mechanical isolation across a gap with high optical transmission.

\bibliography{EPCC}

\section*{ACKNOWLEDGMENTS}
\textbf{Funding:} We gratefully acknowledge financial support from the Villum Foundation Young Investigator Program (Grant No.\ 13170), the Danish National Research Foundation (Grant No.\ DNRF147 - NanoPhoton), Innovation Fund Denmark (Grant No.\ 0175-00022 - NEXUS and Grant no.\ 2054-00008 - SCALE), Independent Research Fund Denmark (Grant No.\ 0135-00315 - VAFL), the European Union’s Horizon Europe research and innovation program (grant no.\ 101098961 - NEUROPIC), and the European Research Council (grant no.\ 101045396 - SPOTLIGHT). \textbf{Author contributions:} B.V.L., M.A., and S.S. conceived the concepts. S.S. supervised the project. R.E.C., B.V.L., and C.A.R. designed the electronic-photonic circuit crossings and performed the numerical calculations. B.V.L., M.A., M.T.S., and S.S. designed the chips. M.A. and C.A.R. fabricated the samples. B.V.L. and M.A. carried out the experiments. B.V.L., M.A., C.A.R., K.T., and S.S. contributed to theory and data analysis. B.V.L. and S.S. wrote the manuscript with contributions and input from all authors. \textbf{Competing interests:} M. A., B. V. L., S. S., and R.E.C. declare a relevant patent (patent number: US11726262B2). The remaining authors declare no competing interests. \textbf{Data and materials availability:} All data are available in the manuscript or the supplementary materials.
\end{document}



\baselineskip24pt


\maketitle

\beginsupplement

\newpage

\section*{1. Intuitive design of electronic-photonic circuit crossings with a \SI{100}{nm} gap}
\begin{figure*}
\centering
\includegraphics[width=0.8\linewidth]{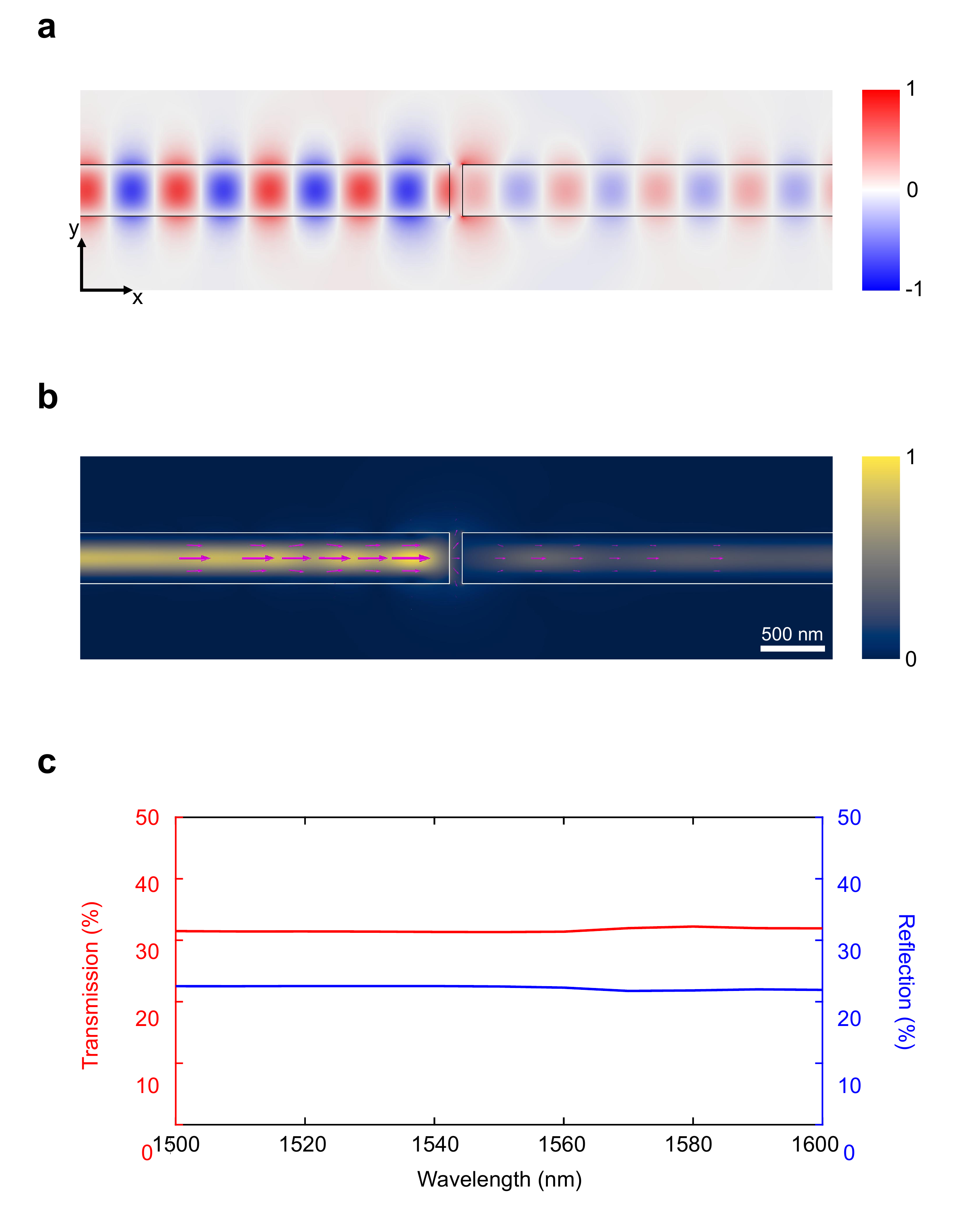}
\caption{\label{fig:naiveNUM}{\textbf{Electronic-photonic circut crossing without optimization.} \textbf{(a)} Normalized transverse electric (TE) field ($E_\text{y}$) distribution across a \SI{100}{nm} isolation gap introduced through a single mode suspended waveguide at \SI{1550}{nm}. \textbf{(b)} Magnitude and direction of the time-averaged power flow in the middle plane of the structure crossing the isolation gap at \SI{1550}{nm}. \textbf{(c)} Calculated transmission (red) and reflection (blue) showing a transmission below \SI{32}{\%} and reflection above \SI{22}{\%} across the full spectral range.}}
\end{figure*}

\begin{figure*}
\centering
\includegraphics[width=0.8\linewidth]{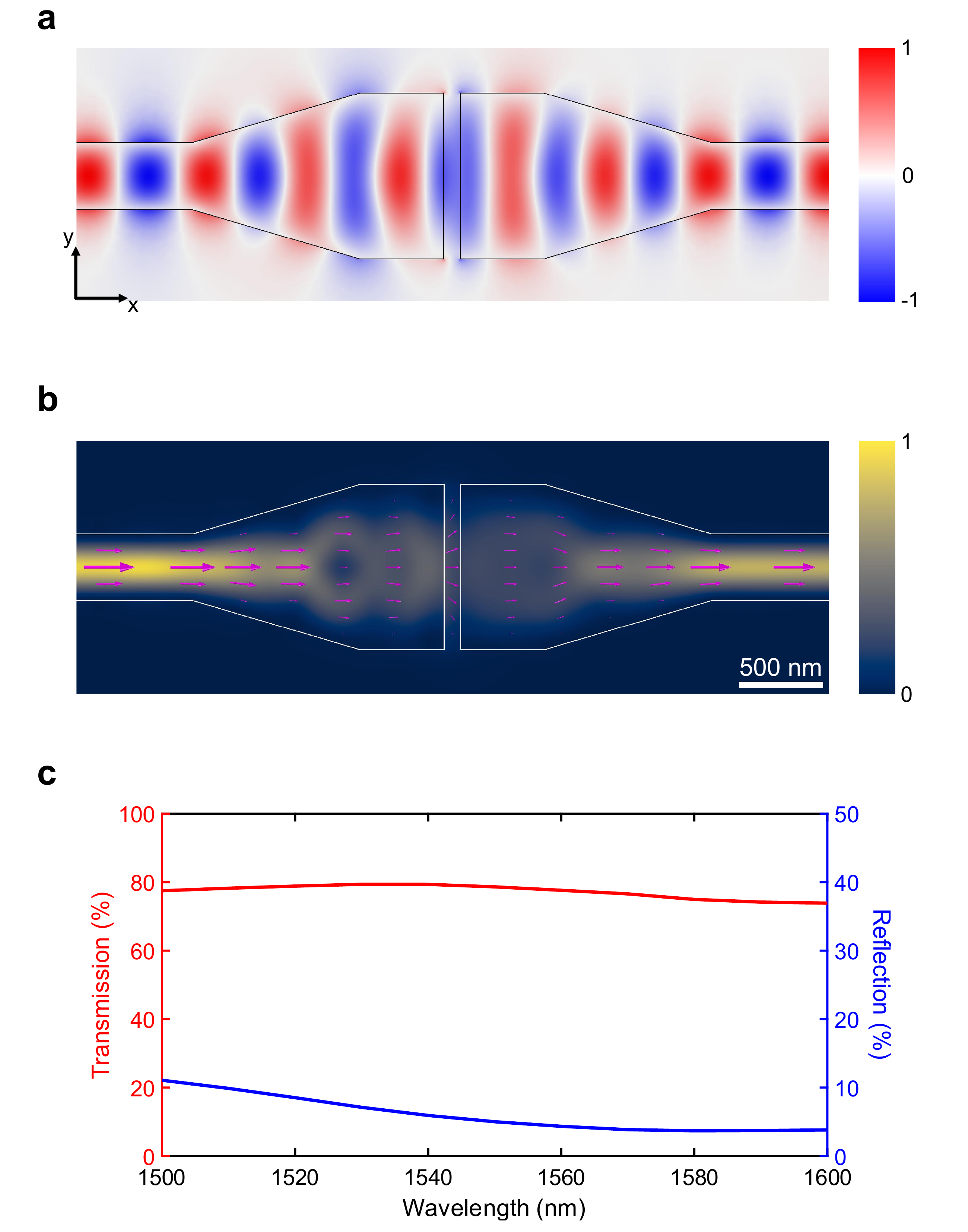}
\caption{\label{fig:TAPNUM}{\textbf{Tapered electronic-photonic circuit crossing with a \SI{100}{nm} gap.} \textbf{(a)} Normalized transverse electric (TE) field ($E_\text{y}$) distribution across a \SI{100}{nm} isolation gap introduced in a tapered single-mode suspended waveguide at \SI{1550}{nm}. \textbf{(b)} Magnitude and direction of the time-averaged power flow in the middle plane of the structure crossing the isolation gap at \SI{1550}{nm}. \textbf{(c)} Calculated transmission (red) and reflection (blue) showing a transmission (below \SI{79}{\%}) and reflection (above \SI{4}{\%}) across the full spectral range.}}
\end{figure*}

Figure~\ref{fig:naiveNUM} illustrates a naive approach to an electronic-photonic circuit crossing (EPCC) made by introducing an isolation air gap in a photonic waveguide. While this approach can provide sufficient electrical isolation, it suffers from poor optical transmission as well as a high reflection. The transmission is below \SI{32}{\%} across the full \SI{100}{nm} bandwidth that we consider in our study, which renders the naive approach practically useless. For example, 20 of such EPCCs in series would lead to a loss of \SI{99}{dB}. To improve the performance of the crossing, one may expand the \SI{400}{nm} suspended single-mode waveguide to a 1 µm wide silicon region through a parameter-optimized tapering section (Fig~\ref{fig:TAPNUM}) which results in a noticeable improvement compared to the naive approach. However, Fig~\ref{fig:TAPNUM} shows that a mode expanding region alone cannot diminish the reflection to sufficiently low values for practical applications. In addition, the \SI{79}{\%} peak power transmission is detrimental to circuits with a few EPCCs. For this simple tapered structure, 20 EPCCs in series would lead to a loss of \SI{20}{dB}.  

\section*{2. Inverse design of electronic-photonic circuit crossings by topology optimization}
\begin{figure*}[t]
\centering
\includegraphics[width=0.85\linewidth]{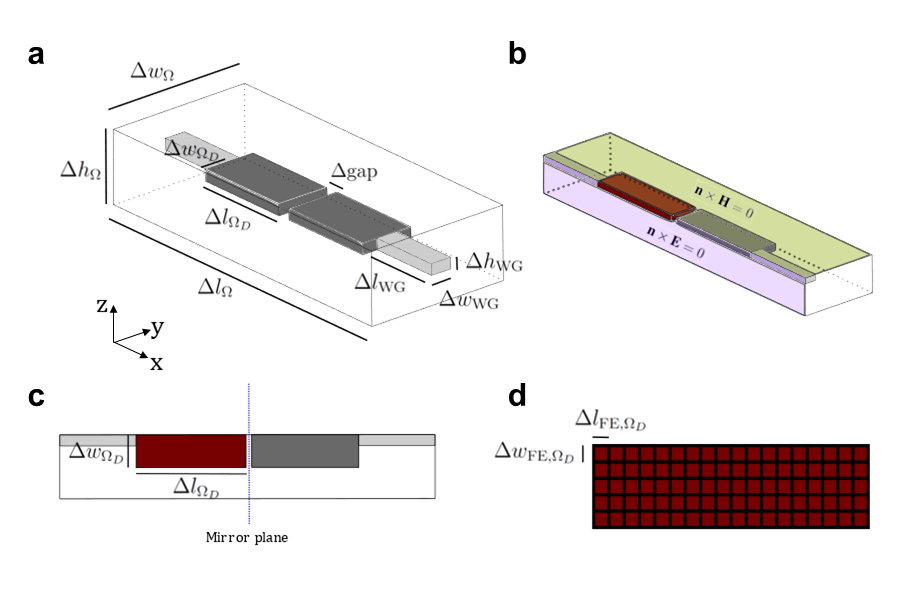}
\caption{\label{fig:Design_domain_topology} {\textbf{Geometry and domain for topology optimization.} \textbf{(a)} Sketch of the EPCC design problem comprised of two identical design domains $\Omega_D$ (dark gray) separated by an air gap, an input waveguide to excite the design domain, and an output waveguide to collect the transmitted light. \textbf{(b)} One-fourth of the model and corresponding boundary conditions exploited to mimic the symmetric nature of the problem. \textbf{(c)} Symmetry along $y$-$z$ plane applied to the design domain and thus reducing the design region (red) to one-eighth. \textbf{(d)} Discretized design region employed to solve the EPCC design problem.}}  
\end{figure*}
Topology optimization \cite{BOOK_TOPOPT_BENDSOE} is an inverse design method, applicable across many areas of physics and engineering \cite{BENDSOE_KIKUCHI_1988,CHRISTIANSEN_SMO_2015,ALEXANDERSEN_ET_AL_2016,MADSEN_2020,JENSEN_SIGMUND_2011,MOLESKY_2018}. When applying topology optimization to solve a structural design problem, the problem is recast as a continuous constrained optimization problem, where the quality of the structure under design is measured using a figure of merit, $\Phi$, which is optimized by changing the structural geometry iteratively and systematically. The structure is represented by a continuous design field, used to interpolate between the material(s) constituting it and its surroundings. The continuity of the design field and all operations applied to it enable the use of computationally efficient adjoint sensitivity analysis \cite{TORTORELLI_ET_AL_1994} to compute gradients of $\Phi$ with respect to the design field, making it possible to use gradient-based optimization algorithms \cite{SVANBERG_2002} to solve the optimization problem. Gradient-based methods are orders of magnitude more efficient than non-gradient-based methods, such as genetic algorithms \cite{GOLDBERG_1989}. Reference\cite{CHRISTIANSEN_SIGMUND_MATLAB_2020} provides a comparison of a gradient- and non-gradient-based method, as well as an introduction to gradient-based topology optimization. Furthermore, topology optimization allows designing and optimizing nanophotonic geometries with no restrictions except user-defined constraints to ensure, e.g., mechanical integrity and feasibility of fabrication \cite{Wang:2018,Christiansen:2019,Christiansen:2020,Hammond:2021}.

We use topology optimization to design EPCC-structures as described in the following. The physics is modelled in the domain, $\Omega \subset \mathbb{R}^3$, containing a silicon membrane structure suspended in an air background (see Fig.~\ref{fig:Design_domain_topology}a). The structure consists of an input and an output rectangular waveguide (light gray in Fig.~\ref{fig:Design_domain_topology}a) connected to a region to be designed, $\Omega_D \subset \Omega$ (dark gray in Fig.~\ref{fig:Design_domain_topology}a), with an air gap situated at its center. We model the physics using Maxwell's equations in the frequency domain by assuming time-harmonic field behaviour and consider a single-mode excitation of the input waveguide. The model domain is truncated using first-order absorbing boundary conditions on all exterior boundaries and input (output) port conditions on the input (output) waveguide. The model symmetries along the $x$-$z$ and $x$-$y$ planes are exploited to reduce the model size (Fig.~\ref{fig:Design_domain_topology}b). Further, mirror symmetry is enforced on the design geometry around the center point in the air gap along the $y$-$z$ plane (Fig.~\ref{fig:Design_domain_topology}c). The region to be designed is marked with red in Fig.~\ref{fig:Design_domain_topology}, b and c. Finally, the design is discretized into piecewise constant elements and restricted to in-plane variations to conform to the fabrication process (Fig.~\ref{fig:Design_domain_topology}d). The model problem is discretized and solved using the finite-element method  \cite{BOOK_FEM_EM_JIN} in COMSOL Multiphysics.

As we are interested in maximizing the transmittance, $T$, across the air gap, while minimizing the reflectance, $R$, back into the input waveguide, we consider the following figure of merit, 
\begin{equation}
    \Phi = \frac{T}{R+1},
\end{equation}
where $T$ denotes the transmittance across the EPCC and $R$ denotes the reflectance from the EPCC back into the input waveguide.
To avoid numerical modelling issues and ensure a final binary design, we use a standard density filter, cf.\ Eq.~(\ref{EQN:SMOOTHING_OPTIM_PROBLEM}) with filter radius $r_f$ and a smoothed threshold operation (Eq.~(\ref{EQN:THRESHOLD_OPTIM_PROBLEM})), with threshold sharpness, $\beta$, and threshold level, $\eta$ \cite{LAZAROV_2011,WANG_ET_AL_2011}. We apply a continuation scheme for $\beta$ starting from the initial value $\beta_{\mathrm{ini}}$ and increase this value by the growth factor $\beta_{\mathrm{g}}$ at every iteration until a maximum value of $\beta_{\mathrm{max}}$ is reached.

\begin{figure*}
\centering
\includegraphics[width=0.8\linewidth]{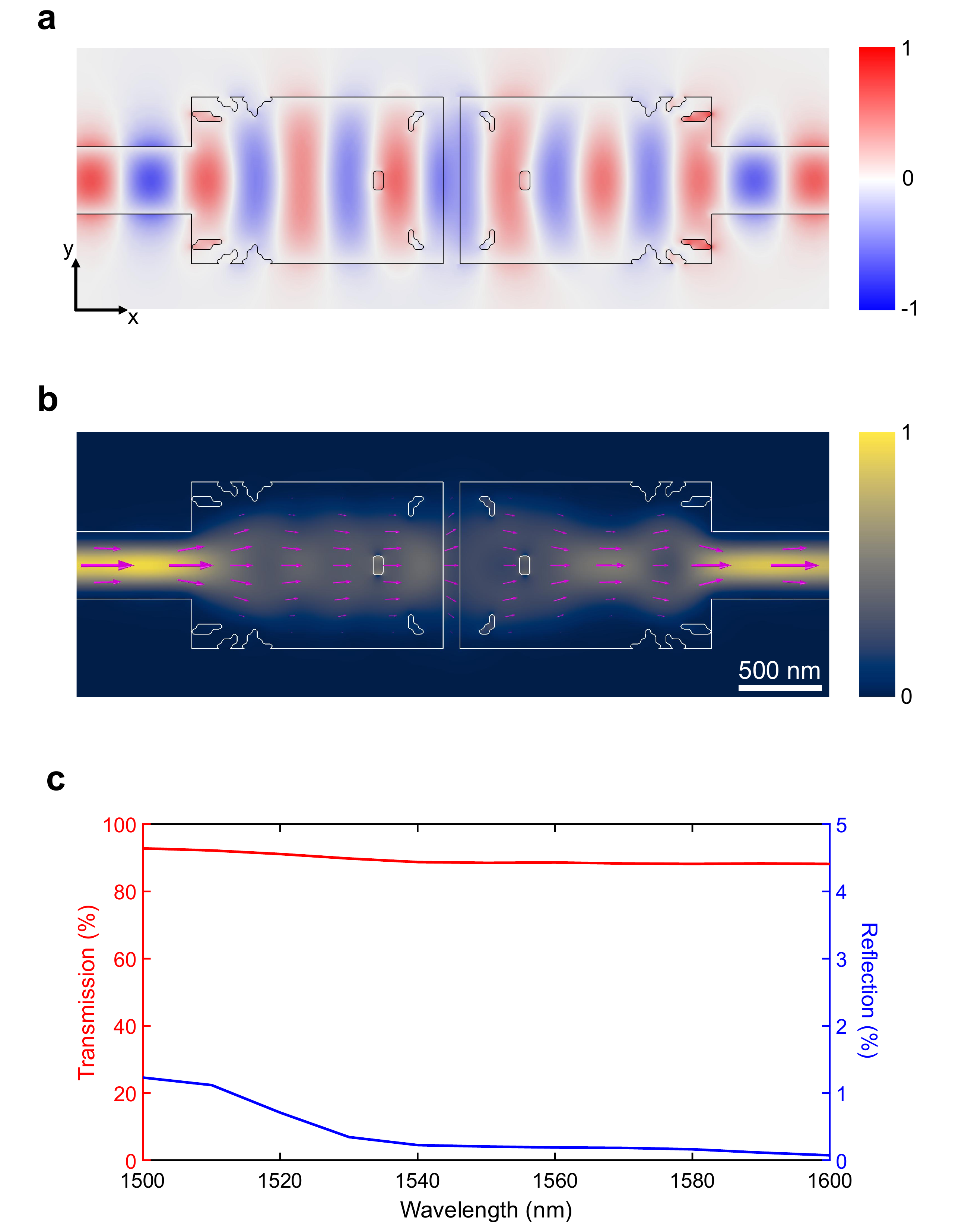}
\caption{\label{fig:TONum}\textbf{Curve-fitted topology-optimized EPCC design.} {\textbf{(a)} Normalized transverse electric (TE) field ($E_\text{y}$) distribution across the EPCC. \textbf{(b)} Magnitude and direction of the time-averaged power flow in the middle plane of the structure for an EPCC constrained to a \SI{100}{nm} isolation gap.\textbf{(c)} Calculated transmission (red) and reflection (blue) of the curve-fitted topology-optimized EPCC over \SI{100}{nm} wavelength span centered at \SI{1550}{nm}.}}
\end{figure*}

In order to enable the design of an EPCC operating across specific bandwidths, e.g., $\lambda \in [\SI{1500}{nm}, \SI{1600}{nm}]$, we formulate a continuous constrained optimization problem, solved to maximize the minimum of $\Phi$ over a set of $\mathcal{N}$ targeted wavelengths. This problem can be written as
\begin{eqnarray}
\underset{\xi(\textbf{r})}{\max} \ \ \underset{\lambda_i}{\min}  &\Phi(\textbf{E}(\lambda_i,\bar{\tilde{\xi}}))& \label{EQN:FOM_OPTIM_PROBLEM} \\
\mathrm{s.t.} &\mathcal{L}_{EM}\left(n(\bar{\tilde{\xi}}(\textbf{r})),\textbf{E}(\lambda_i)\right)= \textbf{f}_i, \ \ \forall \ \ i \in \lbrace 1,2,...,\mathcal{N} \rbrace& \label{EQN:STATE_OPTIM_PROBLEM} \\
&n(\bar{\tilde{\xi}}) = n_{\mathrm{Air}} + \bar{\tilde{\xi}} \left(n_{\mathrm{Si}} - n_{\mathrm{Air}} \right), &  \label{EQN:MATERIAL_INTERPOLATION_OPTIM_PROBLEM} \\
&\bar{\tilde{\xi}} = \frac{\tanh(\beta \cdot \eta) + \tanh(\beta \cdot (\tilde{\xi} - \eta))}{\tanh(\beta \cdot \eta) + \tanh(\beta \cdot (1 - \eta))}, & \label{EQN:THRESHOLD_OPTIM_PROBLEM} \\
&-\left(\frac{r_f}{2\sqrt{3}}\right)^2 \nabla \tilde{\xi}(\textbf{r})+\tilde{\xi}(\textbf{r})=\xi(\textbf{r}),& \label{EQN:SMOOTHING_OPTIM_PROBLEM}  \\
&0 \leq \xi(\textbf{r}) \leq 1.& \label{EQN:DESIGN_FIELD_OPTIM_PROBLEM} 
\end{eqnarray}
Here, $\textbf{r}$ denotes the spatial coordinates, $\xi$ the design field, $\mathcal{L}_{EM}\left(\varepsilon_r(\bar{\tilde{\xi}}(\textbf{r})),\textbf{E}\right)= \textbf{f}$ the Maxwell equations, $n$ the refractive index, and $\textbf{E}$ the electric field. 
The optimization problem is solved using the globally convergent method of moving asymptotes \cite{SVANBERG_2002}. 
The final binary pixel-by-pixel design, obtained by solving the optimization problem, is converted to a set of closed curves, describing the interfaces between silicon and air. The post-processed design is used to compute the numerical performance of the device (see Fig.~\ref{fig:TONum}) and serves as the blueprint for the fabricated devices. Table~\ref{tab:TO parameters} lists the parameters of materials, geometry, finite-element-discretization, and optimization model used to solve the EPCC optimization design problem.

\begin{table}[h!]
  \begin{center}
    \caption{\textbf{Parameters used in the topology optimization of the EPCC.}}
    \label{tab:TO parameters}
    \begin{tabular}{l|c|c} 
      \textbf{Parameter} & \textbf{Variable} & \textbf{Value}\\
      \hline
      Targeted wavelength 1 & $\lambda_1$ & \SI{1500}{nm}\\
      Targeted wavelength 2 & $\lambda_2$ & \SI{1600}{nm}\\
      Refractive index of silicon & $n_{\textbf{S}i}$ & $3.48$\\
      Refractive index of air & $n_{Air}$ & $1.0$\\
      Model domain width & $\Delta w_{\Omega}$ & \SI{1600}{nm}\\
      Model domain height & $\Delta h_{\Omega}$ & \SI{1200}{nm}\\
      Model domain length & $\Delta l_{\Omega}$ & \SI{6000}{nm}\\
      Waveguide width & $\Delta w_{\mathrm{WG}}$ & \SI{400}{nm}\\
      Waveguide height & $\Delta h_{\mathrm{WG}}$ & \SI{250}{nm}\\
      Waveguide length & $\Delta l_{\mathrm{WG}}$ & \SI{1450}{nm}\\
      Design domain width & $\Delta w_{\Omega_D}$ & \SI{500}{nm}\\
      Design domain height & $\Delta h_{\Omega_D}$ & \SI{250}{nm}\\
      Design domain length & $\Delta l_{\Omega_D}$ & \SI{1500}{nm}\\
      Gap width & $\Delta \mathrm{gap}$ & \SI{100}{nm}\\
      Design domain brick element width  & $\Delta w_{\mathrm{FE},\Omega_D}$ & \SI{30}{nm}\\
      Design domain brick element height & $\Delta h_{\mathrm{FE},\Omega_D}$ & \SI{50}{nm}\\
      Design domain brick element length & $\Delta l_{\mathrm{FE},\Omega_D}$ & \SI{30}{nm}\\
      Initial guess for the design field & $\xi_{\mathrm{ini}}$ & $0.1$\\
      Filter radius & $r_f$ & \SI{60}{nm}\\
      Thresholding level & $\eta$ & $0.5$\\
      Initial thresholding sharpness & $\beta_{\mathrm{ini}}$ & $1.0$\\
      Maximum thresholding sharpness & $\beta_{\mathrm{max}}$ & $64.0$\\
      Increment factor & $\beta_{\mathrm{g}}$ & $2.0$\\
      Number of design iterations between each $\beta$-increase & $\mathrm{itr}$ & $50$\\
    \end{tabular}
  \end{center}
\end{table}

We design an EPCC optimized for operating at $\lambda_1 = \SI{1500}{nm}$ and $\lambda_2 = \SI{1600}{nm}$, i.e. targeting the bandwidth $\lambda \in [ \SI{1500}{nm}, \SI{1600}{nm} ]$, by solving the optimization problem detailed above. The post-processed design, converted to a set of closed curves, is shown in Fig.~\ref{fig:TONum}. Figure~\ref{fig:TONum}, a and b, illustrate the max-normalized TE-field ($E_\text{y}$) distribution and the Magnitude and direction of the time-averaged power flow in the middle plane of the structure, across a \SI{100}{nm} isolation gap at \SI{1550}{nm}, respectively. Figure~\ref{fig:TONum}c shows the calculated transmission and reflection spectra of the topology-optimized device showing a transmission above $88.2\%$ with reflection smaller than $1.2\%$ over the \SI{100}{nm} wavelength bandwidth centered at \SI{1550}{nm}. The topology optimization is performed for the nominal device-layer thickness of \SI{250}{nm} for the silicon-on-insulator (SOI) wafers. However, we measured the thickness of the device layer using variable-angle spectroscopy ellipsometry after fabrication and found a thickness of \SI{240}{nm}. To better approximate the fabricated device, the calculations presented in Fig.~\ref{fig:TONum} (and in Figs.~1d-f and ~2b in the main text) is carried out for a device-layer thickness of \SI{240}{nm}. We attribute the excellent performance obtained for the slightly different device-layer thickness to an inherent robustness of our design, which is likely due to the broadband optimization procedure.

\begin{figure*}
\centering
\includegraphics[width=0.8\linewidth]{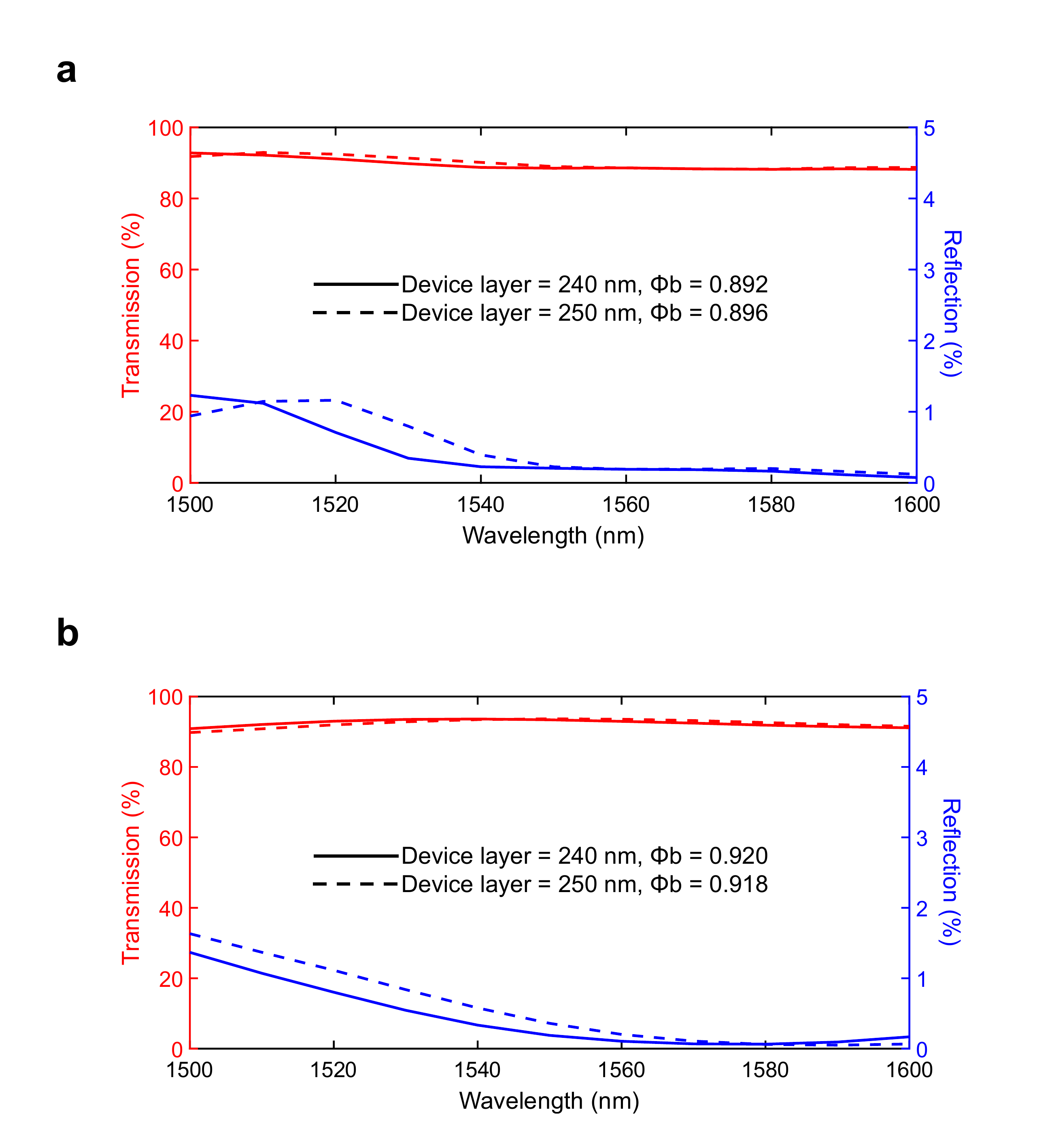}
\caption{\label{fig:TOandHYB}{\textbf{Comparison of curve-fitted and tapered topology-optimized EPCCs}. Calculated optical transmission and reflection of \textbf{(a)} curve-fitted and \textbf{(b)} tapered topology-optimized design for \SI{250}{nm} and \SI{240}{nm} device thicknesses, corresponding to the nominal and measured device layer thickness, respectively.}}
\end{figure*}

By analyzing and simulating the topology-optimized design, we identify that some of the designed features are responsible for expanding the optical mode. Thus, as the final design step in order to simplify the design geometry, we replace these features with a parameter-optimized tapering section as shown in Fig.~ 1. In order to compare the performance of the tapered design with the curve-fitted topology-optimized design, we define a broadband figure of merit ($\Phi_b$) as, 
\begin{equation}
\Phi_b =\frac{1}{N} \sum_{i=1}^{N} \frac{T(\lambda_i)}{R(\lambda_i)+1}
\end{equation}
Where, N denotes the number of wavelengths at which the spectra are calculated, i.e., $11$ (from $1500$ to \SI{1600}{nm} with a \SI{10}{nm} interval) in Figs.~1f and \ref{fig:TONum}c. 
Figure~\ref{fig:TOandHYB}, a and b, shows the transmission and reflection spectra along with the corresponding broadband figure of merit ($\Phi_b$) of the curve-fitted and tapered topology-optimized design calculated for two different device thicknesses: (i) \SI{250}{nm} corresponding to the original thickness of the device layer of the SOI wafer, and (ii) \SI{240}{nm} corresponding to the thickness of the fabricated device.
Here, two main observations can be made: (i) the \SI{10}{nm} thinner device layer does not have a significant influence on the performance of the device and only leads to a small blue-shift in the transmission and reflection spectra showing the robustness of the designed EPCC to small variations in device thickness, and (ii) replacing the mode-expanding features with a tapering section results in an overall better performance, when averaged across the full \SI{100}{nm} bandwidth.

\begin{figure*}
\centering
\includegraphics[width=0.8\linewidth]{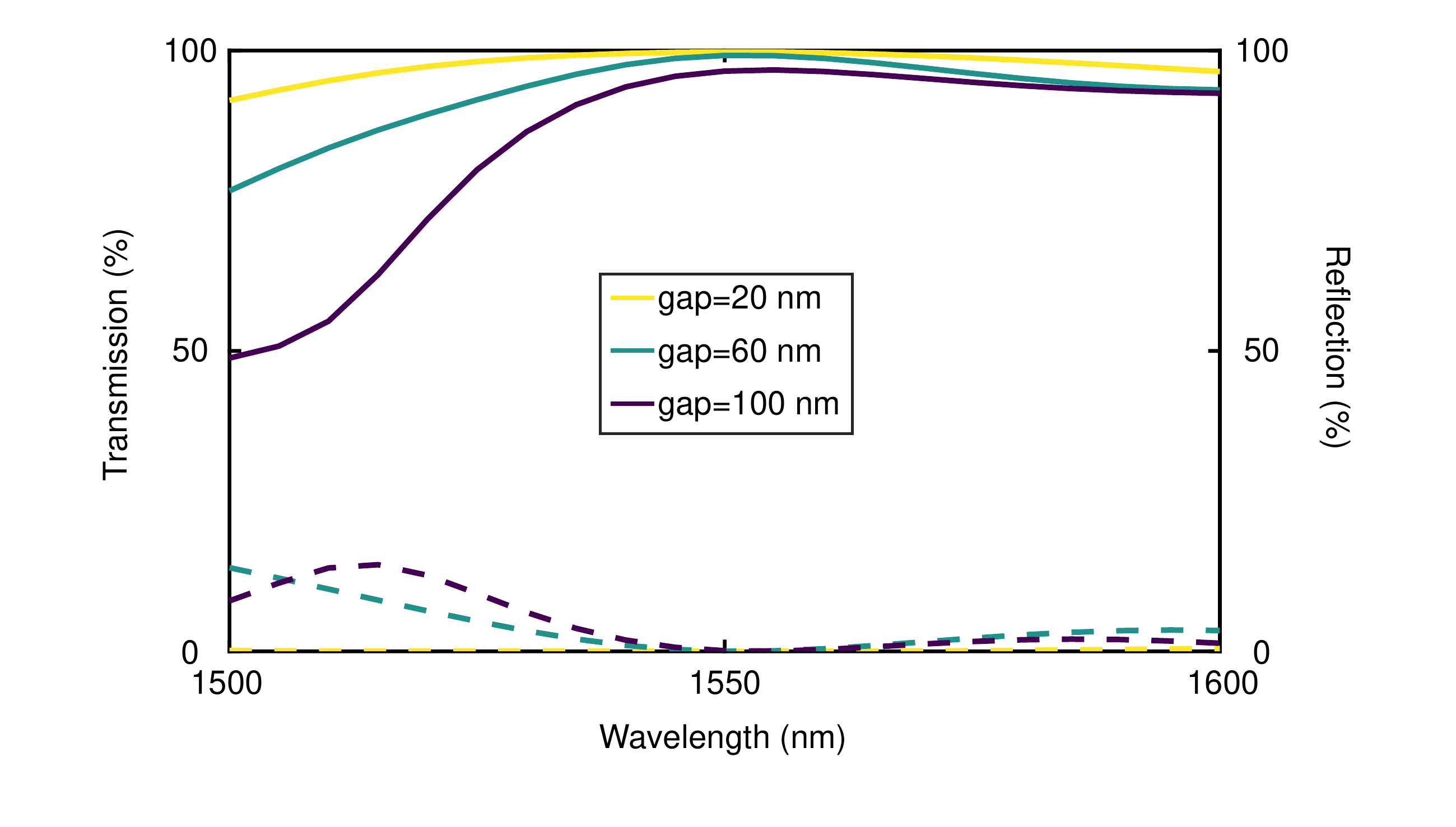}
\caption{\label{fig:UnityEPCC}{\textbf{Topology-optimized EPCCs with near-unity transmission.} Calculated transmission (solid) and reflection (dashed) of the curve-fitted topology-optimized EPCCs for crossings with \SI{20}{nm} (yellow), \SI{60}{nm} (green), and \SI{100}{nm} (purple) isolation gaps optimized for maximum transmission at \SI{1550}{nm} corresponding to the designs shown in Fig.~4 in the main text.}}
\end{figure*}
In addition to the original design presented above and explored experimentally, we design a set of EPCCs for a few different isolation gaps including a very small, yet feasible, gap of \SI{20}{nm} in order to illustrate the generality of our approach for routing optical and electrical signals in large-scale circuits. Encouraged by the already high transmission obtained for the \SI{100}{nm} designs, we aim for reaching transmissions close to unity. Since the reflection is bound by $T+R=1$, we disregard the reflection in the figure of merit and simply define it as $\Phi =T$, where $T$ is power transmission across the isolation trench only at $\lambda=$\SI{1550}{nm}. The curve-fitted topology-optimized design of these EPCCs are presented in Fig.~4 in the main text, and the calculated transmission and reflection are shown in Fig~\ref{fig:UnityEPCC}.

\section*{4. Fabrication of electronic-photonic circuit crossing and nanoelectromechanical switch}
\begin{figure*}
\centering
\includegraphics[width=0.8\linewidth]{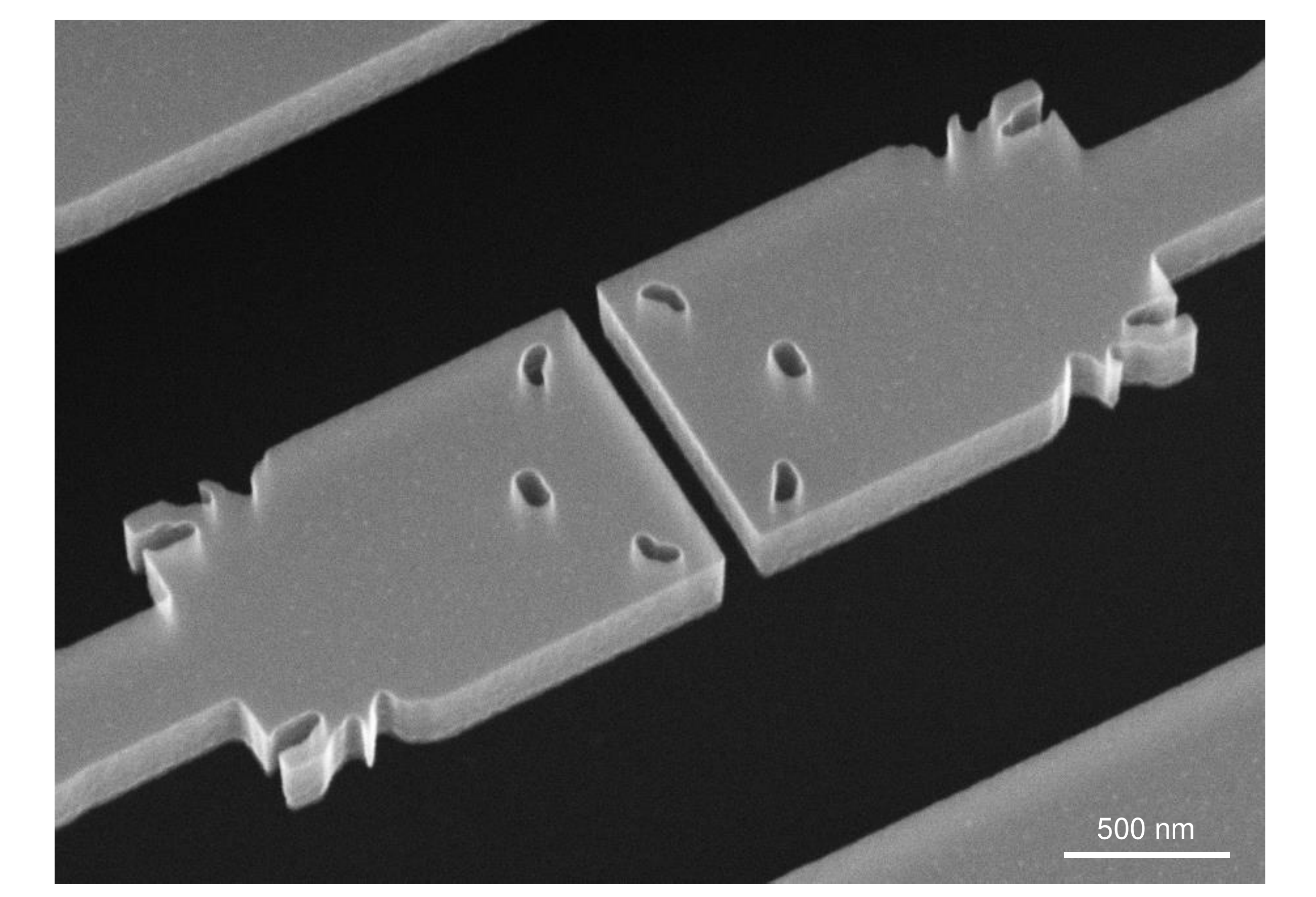}
\caption{\label{fig:TO_Fab}{\textbf{Fabricated curve-fitted topology-optimized EPCC.} A scanning electron micrograph (\SI{30}{\degree} tilted view) of the curve-fitted topology-optimized EPCC.}}
\end{figure*}
The samples are fabricated from undoped (100) \SI{150}{mm} SOI wafers (SOITEC) with a \SI{3}{\micro\meter} buried oxide. The nominally undoped silicon device layer has a resistivity of $\sim\SI{10}{\ohm\centi\meter}$, which is sufficiently low for making electrical contacts while imposing minimal optical absorption losses. The device layer has a nominal thickness of $\SI{250}{nm}\pm\SI{30}{nm}$. We measure the thickness of the device layer to be $\SI{240}{nm}$ using variable-angle spectroscopy ellipsometry (M2000XI-210 ellipsometer, J.A.\ Woollam Co.). More details on the lithography and release processes along with a discussion of the device-layer thickness can be found in Ref.~\cite{Albrechtsen:21}.
The devices are patterned with a single electron-beam lithography step and etched directly from the soft-mask using deep reactive-ion etching (DRIE) in a continuous cryogenic ($-\SI{19}{\celsius}$) inductively-coupled plasma (ICP). Figure~\ref{fig:TO_Fab} shows a scanning electron micrograph of the fabricated curve-fitted topology-optimized EPCC.

A detailed process flow of the fabrication is included below for the cutback and switch network samples.

\textbf{Fabrication flow for sample 1: Cutback circuits}
\begin{enumerate}[noitemsep]
    \item Pattering of devices (electron-beam lithography):
    \begin{enumerate}[noitemsep]
        \item Prepare the sample: A 25-by-\SI{25}{mm} chip is cleaved from a wafer and cleaned by sequential submersion in acetone, isopropanol (IPA), deionized water (DIW), IPA, and finally, it is dried with dry \ch{N2}.
        \item Spin-coating: The chip is fixed to a \SI{100}{mm} silicon carrier wafer using heat-sensitive wax (Crystalbond 555, 821-2) and \SI{180}{nm} chemically semi-amplified resist (AllResist CSAR6200.09) is spin-coated at \SI{6000}{rpm} (\SI{10000}{rpm\per\second}) for \SI{60}{s} using an automatic spin-coater (Süss MicroTec Gamma 4M cluster). The solvent (anisole) is removed with a \SI{180}{s} softbake at \SI{180}{\celsius}.
        \item Deposition of decharging layer: \SI{20}{nm} aluminium is deposited by thermal evaporation (Kurt J. Lesker Nano 36 Thermal Evaporator) at a rate of \SI{1}{\angstrom\per\second}.
        \item Exposure: The patterns are exposed with a \SI{100}{\kilo\electronvolt} electron-beam lithography system (Jeol JBX-9500FSZ), with a clearing dose of \SI{180}{\micro\coulomb\per\cm\squared} and a current of $I=\SI{196.8}{pA}$.
        \item Development: The aluminium decharging layer is removed by submersion in tetramethylammonium hydroxide (TMAH, \SI{2.38}{\percent} in water, AZ 726 MIF) and the chip is cleaned in DIW. The e-beam resist is developed for \SI{60}{s} in amyl acetate (AR 600-546) and the chip is cleaned in IPA and dried with dry \ch{N2}.
    \end{enumerate}
    \item Pattern transfer (ICP-DRIE):
    \begin{enumerate}
    	\item Etching: The chip is fixed to a \SI{100}{mm} \ch{SiO2}-covered silicon carrier wafer using heat-sensitive wax (Crystalbond 555, 821-2) and etched at \SI{-19}{\celsius} by DRIE (SPTS Pegasus), using a continuous flow of \SI{75}{sccm} \ch{C4F8} and \SI{38}{sccm} \ch{SF6}. The plasma has a pressure of \SI{3}{mTorr} and is powered by a \SI{800}{W} ICP-power and \SI{40}{W} platen power. The first \SI{3}{s} has an increased pressure of \SI{15}{mTorr} to strike the plasma, and the total etching time is \SI{80}{s}.
    	\item Resist strip: The resist is stripped by submersion in dioxolane (AR 600-71), cleaned in IPA, and dried in dry \ch{N2}.
    	\item Plasma ashing: Remaining organic components and fluorocarbon residues from the dry-etch are cleaned for \SI{10}{min} in a barrel asher (TePla, 300 Semi Auto Plasma Processor) with \SI{400}{sccm} \ch{O2}, \SI{70}{sccm} \ch{N2}, \SI{1.23}{mbar} pressure, and \SI{1000}{W} power.
    \end{enumerate}
    \item Wet release etch of suspended silicon structures:
    \begin{enumerate}
    	\item Etching of buried oxide (\ch{SiO2}): The sample is submerged in buffered hydrofluoric acid (\SI{5}{\percent} in water, etch rate \SI{100}{nm\per\min}) for \SI{45}{min}.
    	\item Cleaning: The sample is submerged for \SI{5}{min} in running DIW.
    	\item Drying: The sample is placed in a closed chamber for \SI{10}{min} with heated ethanol to exchange DIW to ethanol, which has lower surface tension. The chip is left to dry at room temperature.
    \end{enumerate}
\end{enumerate}

\textbf{Fabrication flow for sample 2: Switch network}

Patterning of the silicon device layer is carried out following steps 1-2 of the cutback circuits. Metallization and release of fragile suspended silicon membranes are carried out using the following two fabrication steps.
\begin{enumerate}
	\item Metallization and definition of bonding pads:
	\begin{enumerate}
		\item Spin-coating: The chip is fixed to a \SI{100}{mm} silicon carrier wafer using heat-sensitive wax (Crystalbond 555, 821-2) and \SI{2}{\micro\meter} negative i-line (\SI{365}{nm}) resist (AZ nLof 2020) is spin-coated at \SI{3300}{rpm} (\SI{10000}{rpm\per\second}) for \SI{60}{s} and softbaked for \SI{60}{s} at \SI{90}{\celsius} using an automatic spin-coater (Süss MicroTec Gamma 4M cluster).
		\item Lithography: The sample is exposed (\SI{169}{mJ\per\centi\meter}) using maskless UV-lithography (Heidelberg, MLA100) without removing the chip from the carrier wafer.
		\item Development: A post-exposure bake (\SI{60}{s}, \SI{110}{\celsius}) is carried out and the sample is developed tetramethylammonium hydroxide (TMAH, \SI{2.38}{\percent} in water, AZ 726 MIF) for \SI{60}{s} using an automatic developer (Süss MicroTec Gamma 2M developer). Lastly, the chip is detached from the carrier wafer, cleaned in DIW, and dried with dry \ch{N2}.
		\item Metal deposition: \SI{5}{nm} chromium is deposited (rate: \SI{5}{\angstrom\per\second}) to ensure adhesion followed by \SI{20}{nm} gold (rate: \SI{2}{\angstrom\per\second}) for good electrical contact. Both depositions are done with e-beam evaporation (Ferrotec Temescal FC-2000) from a base pressure of $10^{-6}\SI{}{mTorr}$.
		\item Lift-off: The sample is submerged in N-Methyl-2-pyrrolidone (NMP, 1165 Remover) until the lift-off is complete (\SI{4}{\hour}), and the sample is cleaned in IPA and dried with dry \ch{N2}.
	\end{enumerate}
	\item Release etch of silicon membranes:
	\begin{enumerate}
	\item Cleaning: The sample is cleaned by sequential submersion into NMP, acetone, and IPA, and dried with dry \ch{N2}. It is then cleaned in a barrel asher (TePla, 300 Semi Auto Plasma Processor) with \SI{400}{sccm} \ch{O2}, \SI{70}{sccm} \ch{N2}, \SI{1.13}{mbar} pressure, and \SI{200}{W} power for \SI{2}{min}.
	\item Pre-release bake: The sample is dehydrated and decarbonated by baking it on a \SI{200}{\celsius} hotplate for \SI{5}{min}.
	\item Release etch: The sample is placed in a single-wafer Primaxx uEtch chamber (SPTS) and the buried oxide is etched \SI{4.5}{\micro\meter} over 3 cycles to minimize stiction. The first is \SI{10}{min} with \SI{1000}{sccm} \ch{N2}, \SI{400}{sccm} ethanol, and \SI{525}{sccm} anhydrous hydrofluoric-acid (\SI{99.995}{\percent}), resulting in a mean pressure of \SI{132.2}{Torr} and an etch rate of \SI{150}{nm\per\min} (total \SI{1.5}{\micro\meter} etched). The two next cycles are identical, each \SI{24}{min} with \SI{1250}{sccm} \ch{N2}, \SI{350}{sccm} ethanol, and \SI{310}{sccm} \ch{HF}, resulting in a mean pressure of \SI{132.7}{Torr} and an etch rate of \SI{63}{nm\per\min} (total \SI{3}{\micro\meter} etched).
	\end{enumerate}
\end{enumerate}

\section*{5. Experimental procedure}
\begin{figure*}
\centering
\includegraphics[width=0.8\linewidth]{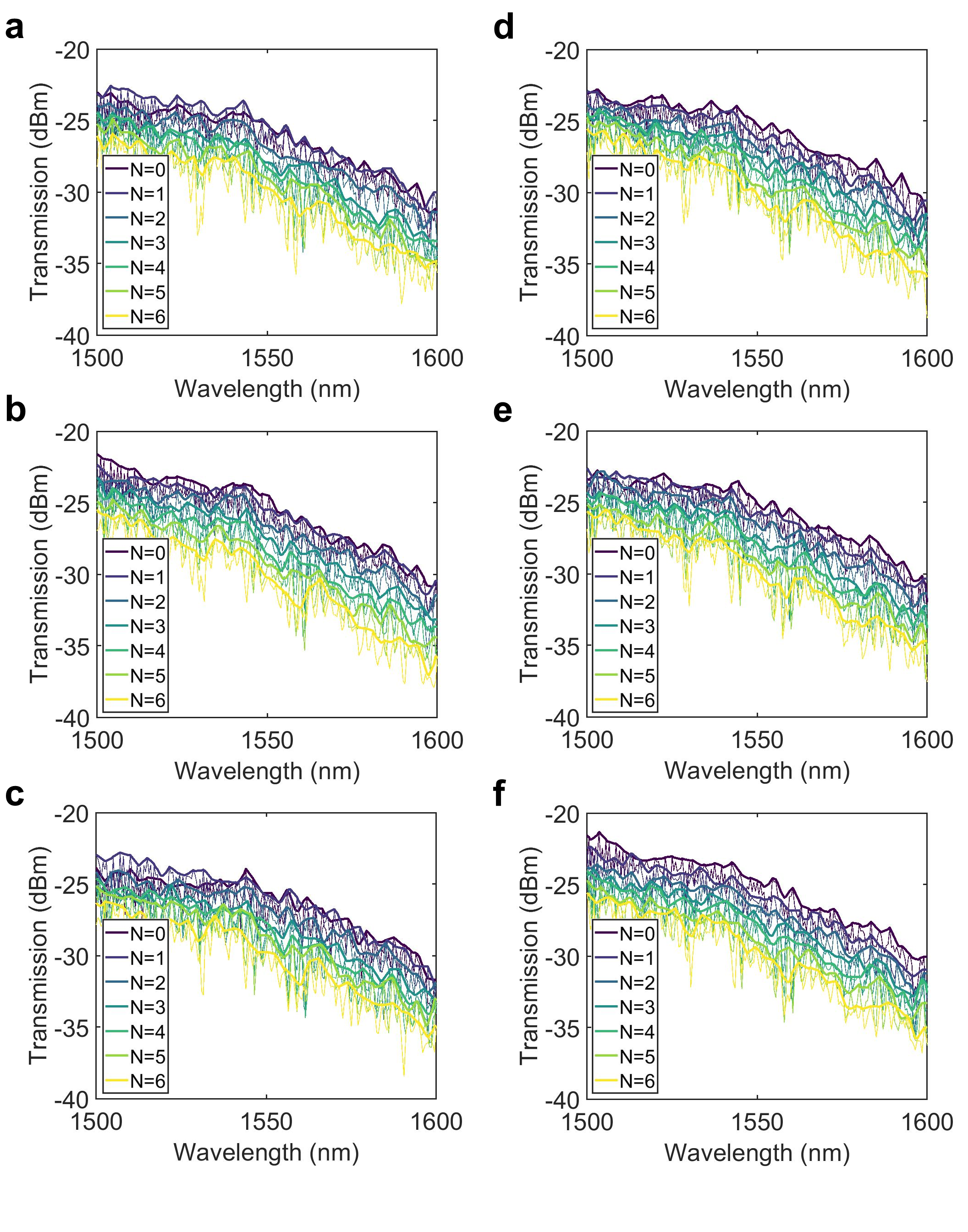}
\caption{\label{fig:RawDataTO}{\textbf{Characterization of curve-fitted topology-optimized EPCC.} \textbf{(a-f)} Raw data and envelopes (bold lines) for 6 identical sets of photonic circuits each comprised of 7 transmission lines with 0 to 6 curve-fitted topology-optimized EPCCs.}}
\end{figure*}

\begin{figure*}
\centering
\includegraphics[width=0.8\linewidth]{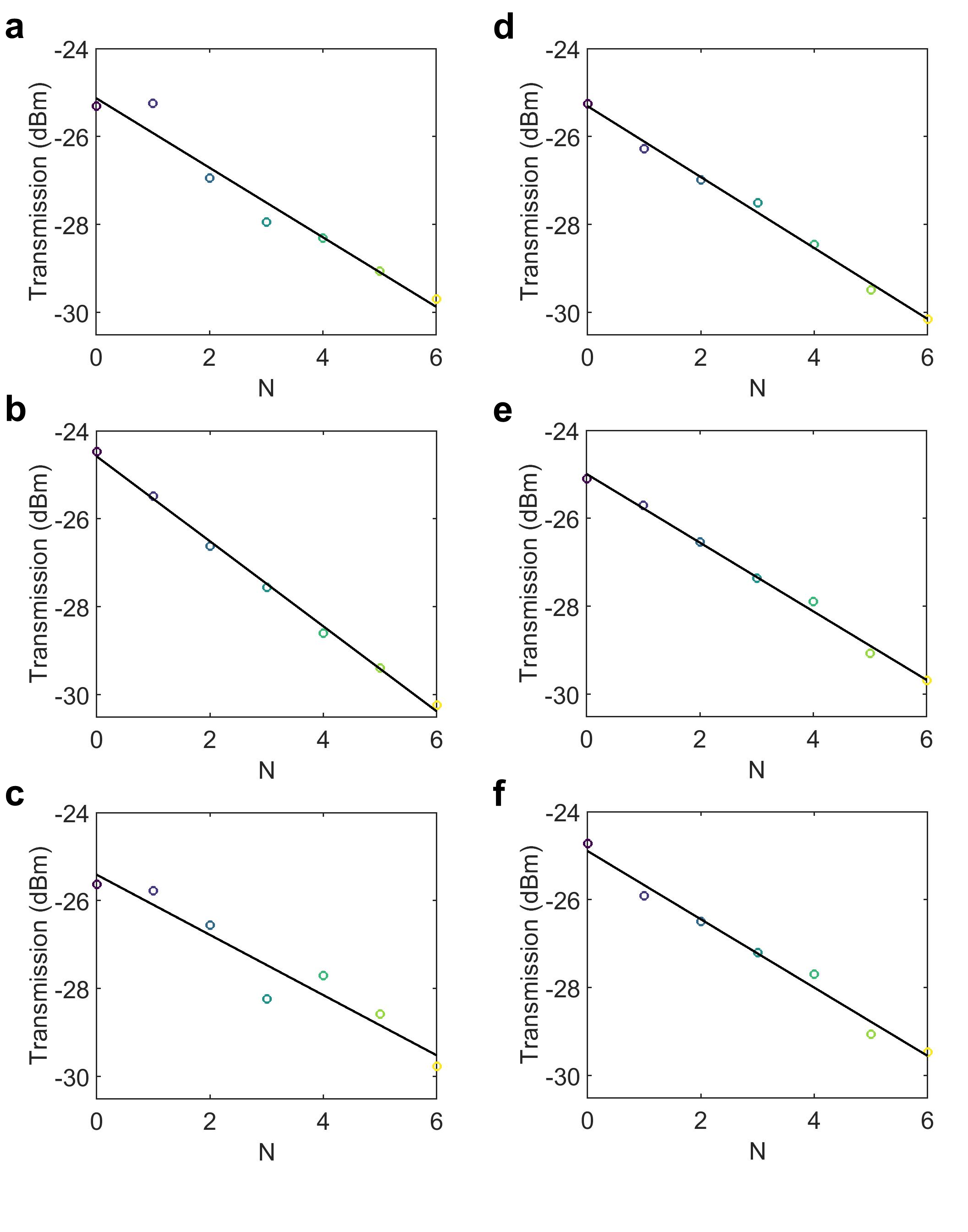}
\caption{\label{fig:CutBackTO}{\textbf{Assessment of transmission efficiency of curved-fitted topology-optimized EPCC through the cut-back method.} \textbf{(a-f)} Transmission of each set of circuits vs number of EPCCs at \SI{1550}{nm}. The slope of the linear fit (black line) gives the propagation loss and thus the transmission efficiency.}}
\end{figure*}

Loss measurements of each device can be accurately performed by the cutback method, where the device loss is extracted by comparing the optical transmission through a varying number of devices, in this case $N= \{0,1,2,3,4,5,6\}$ in an optical transmission-line circuit as shown in Fig.~2a in the main text. A free-space optical setup is used to carry out transmission measurements on each circuit. A linearly polarized tunable laser (Santec TSL-710) is coupled to the chip via an input grating coupler while swept from $1500$ to \SI{1600}{nm} with a step of \SI{0.5}{nm}. A grating coupler with 90-degree rotation, and hence orthogonal polarization, is used to couple the output light to a single-mode optical fiber, where the collected light is delivered to an optical spectrum analyzer (Yokogawa AQ6370D) for the spectral measurements. The orthogonal in- and output grating couplers eliminate the direct laser light reflection from the chip so that only light transmitted through the device is recorded.

Figure~\ref{fig:RawDataTO} shows raw transmission data of six nominally identical sets of photonic circuits in which each set is comprised of 7 transmission lines with 0 to 6 curve-fitted topology-optimized devices. Because of reflection at the location of the grating couplers, a Fabry-Perot cavity is created at each circuit that results in fringes in the transmission spectra. The effect of the Fabry-Perot cavity can be discarded by considering the upper peak envelope of the transmission spectra (bold lines in Fig~\ref{fig:RawDataTO}, a-f), which is obtained by calculating the local maxima through a moving window of 5 data points (\SI{2}{nm} spectral window) followed by a moving average of the same window size.  

In order to evaluate the experimental transmission of the curve-fitted topology-optimized device, a linear fit is applied to each set of data associated with each nominally identical set of photonic circuit at each collected data point, i.e, each wavelength of the obtained transmission. The transmission loss, $L$, can thereby be determined by directly measuring the gradient of the fit, and thus the transmission efficiency of the device can be measured by $T=10^\frac{-L}{10}$.   

\begin{figure*}
\centering
\includegraphics[width=0.8\linewidth]{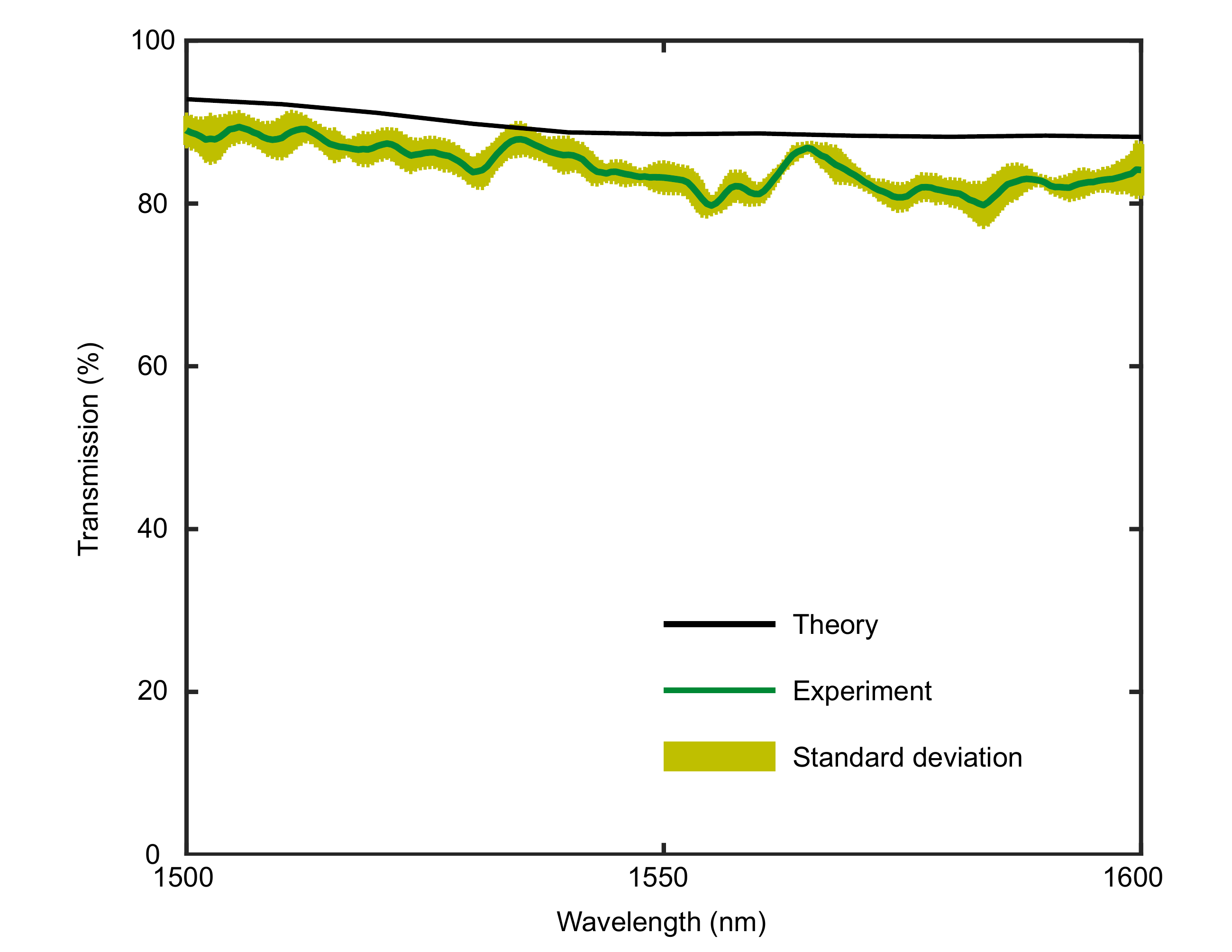}
\caption{\label{fig:TO_exp}{\textbf{Measured optical transmission of curve-fitted topology-optimized EPCC obtained through the cut-back method vs theory}. The transmission spectrum is extracted by evaluating the transmission efficiency through the method shown in Fig.~\ref{fig:CutBackTO} at each data points corresponding to wavelengths of the tunable monochromatic light source used to conduct the experiment. The green shaded area represents the standard deviation from the mean transmission of the six replicas. Numerical calculations were performed for the measured device-layer thickness of \SI{240}{nm}.}}
\end{figure*}

\begin{figure*}
\centering
\includegraphics[width=0.8\linewidth]{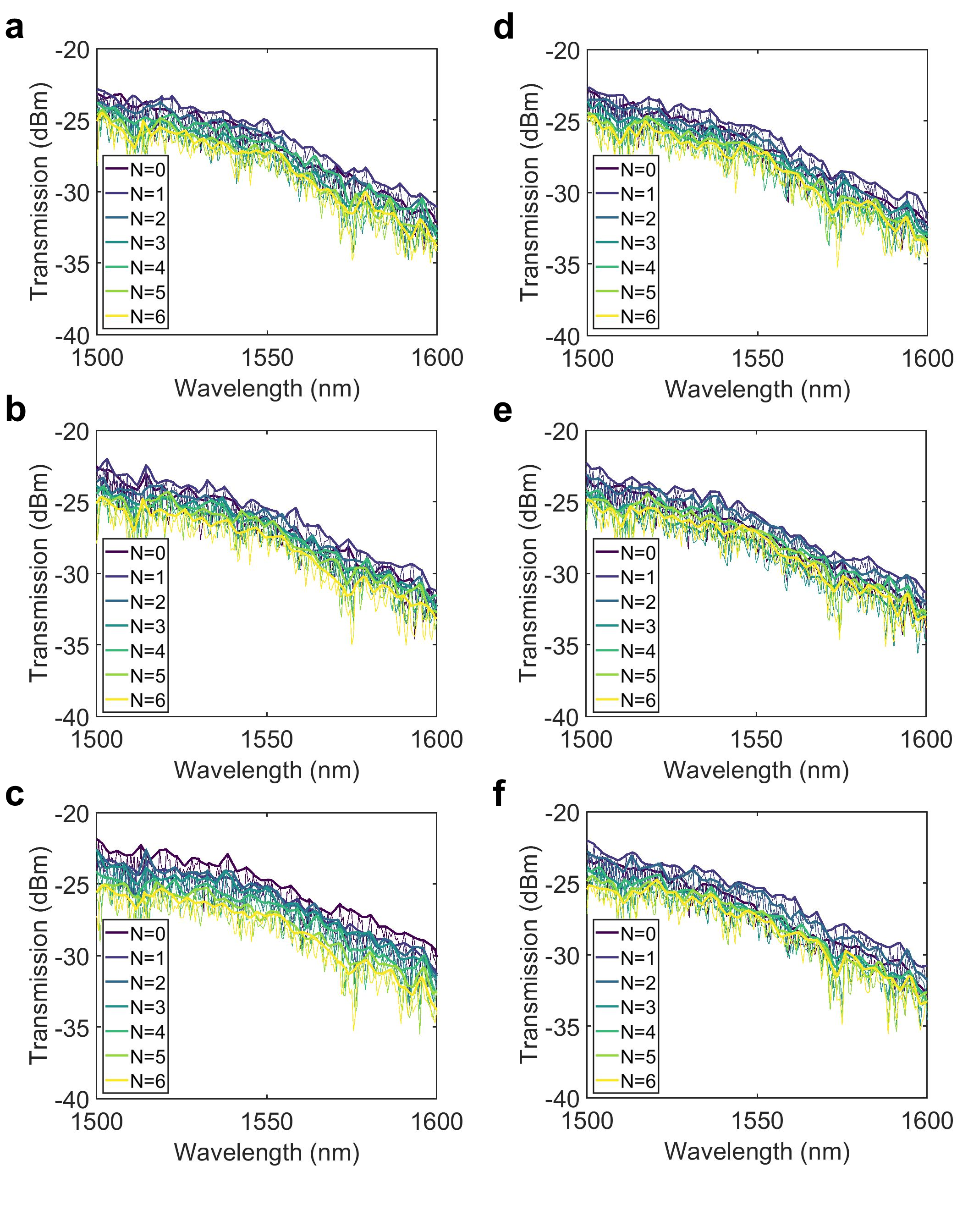}
\caption{\label{fig:RawDataHYB}{\textbf{Characterization of tapered topology-optimized EPCC.} \textbf{(a-f)} Raw data and envelopes (bold lines) for 6 identical sets of photonic circuits each comprised of 7 transmission lines with 0 to 6 tapered topology-optimized EPCCs.}}
\end{figure*}

\begin{figure*}
\centering
\includegraphics[width=0.8\linewidth]{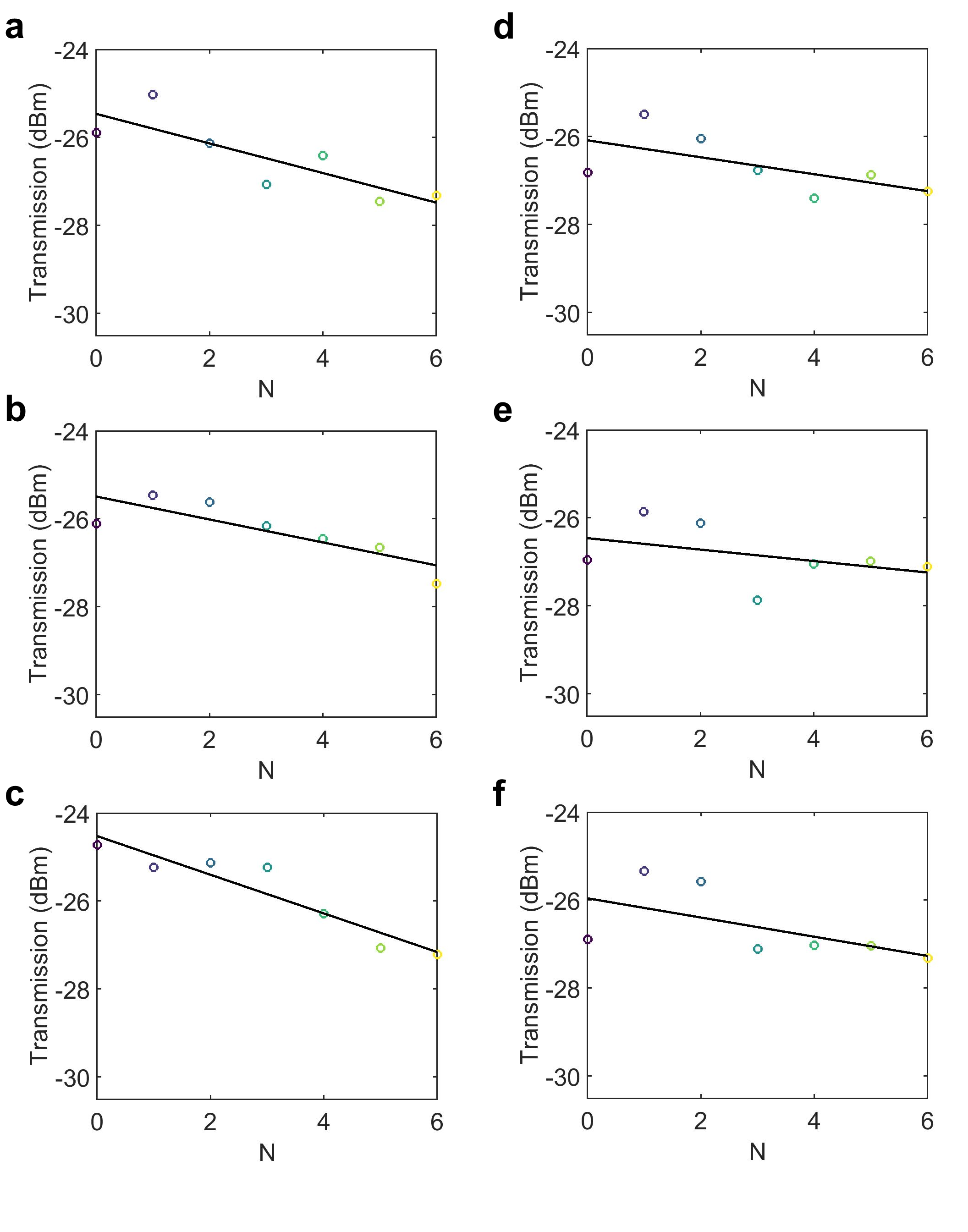}
\caption{\label{fig:CutBackHYB}\textbf{Assessment of transmission efficiency of tapered topology-optimized EPCC through the cut-back method.} {\textbf{(a-f)} Transmission of each set of circuits vs number of EPCCs at \SI{1550}{nm}. The slope of the linear fit (black line) gives the propagation loss and thus the transmission efficiency.}}
\end{figure*}
Figure~\ref{fig:CutBackTO}, a-f, shows this process at $\lambda=$\SI{1550}{nm}. Similarly, the transmission spectrum of the device can be extracted from $1500$ to \SI{1600}{nm} for each set of photonic circuits that in total provides 6 transmission spectra extracted from 6 sets of nominally identical photonic circuits. Figure ~\ref{fig:TO_exp} illustrates the mean transmission (green curve) and the standard deviation from the mean (green shaded area), obtained through conducting transmission measurements on in total 42 transmission line circuits, and compares it with the theoretical calculations for curve-fitted topology-optimized design. A similar approach, as shown in Figs.~\ref{fig:RawDataHYB} and \ref{fig:CutBackHYB}, is used to extract the transmission spectra of the tapered topology-optimized EPCC presented in Fig.~2b of the main text.

\section*{6. In-plane nanoelectromechanical add-drop switch using electronic-photonic circuit crossings}

\begin{figure*}
\centering
\includegraphics[width=0.8\linewidth]{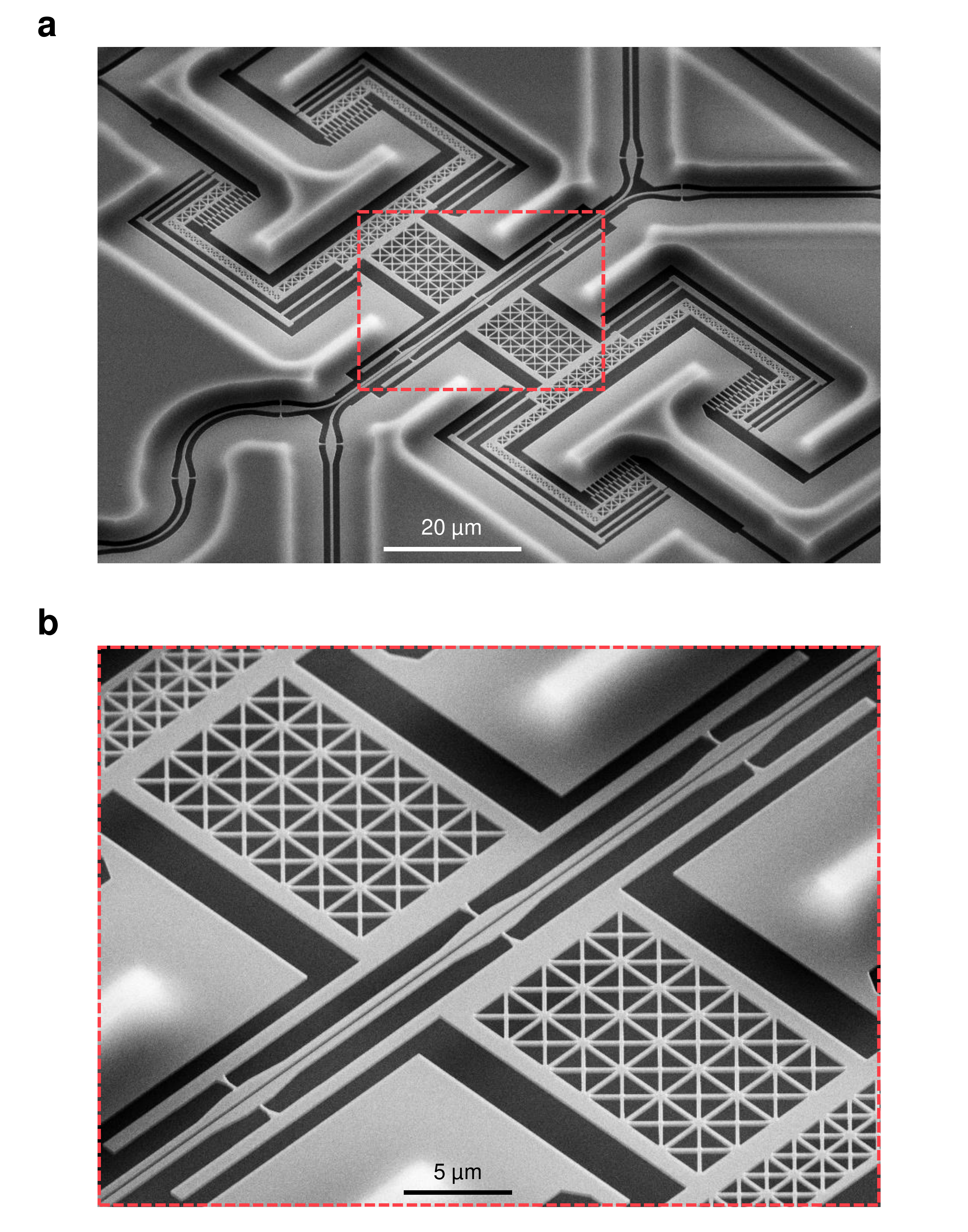}
\caption{\label{fig:directional coupler fab}{\textbf{Fabricated nanoelectromechanical directional coupler.} \textbf{(a)} \SI{45}{\degree} tilted scanning electron micrograph of one switching element comprised of an adjustable directional coupler driven via two comb-drive actuators. \textbf{(b)} A zoomed-in scanning electron micrograph of the directional coupler.}}
\end{figure*}

\begin{figure*}
\centering
\includegraphics[width=0.8\linewidth]{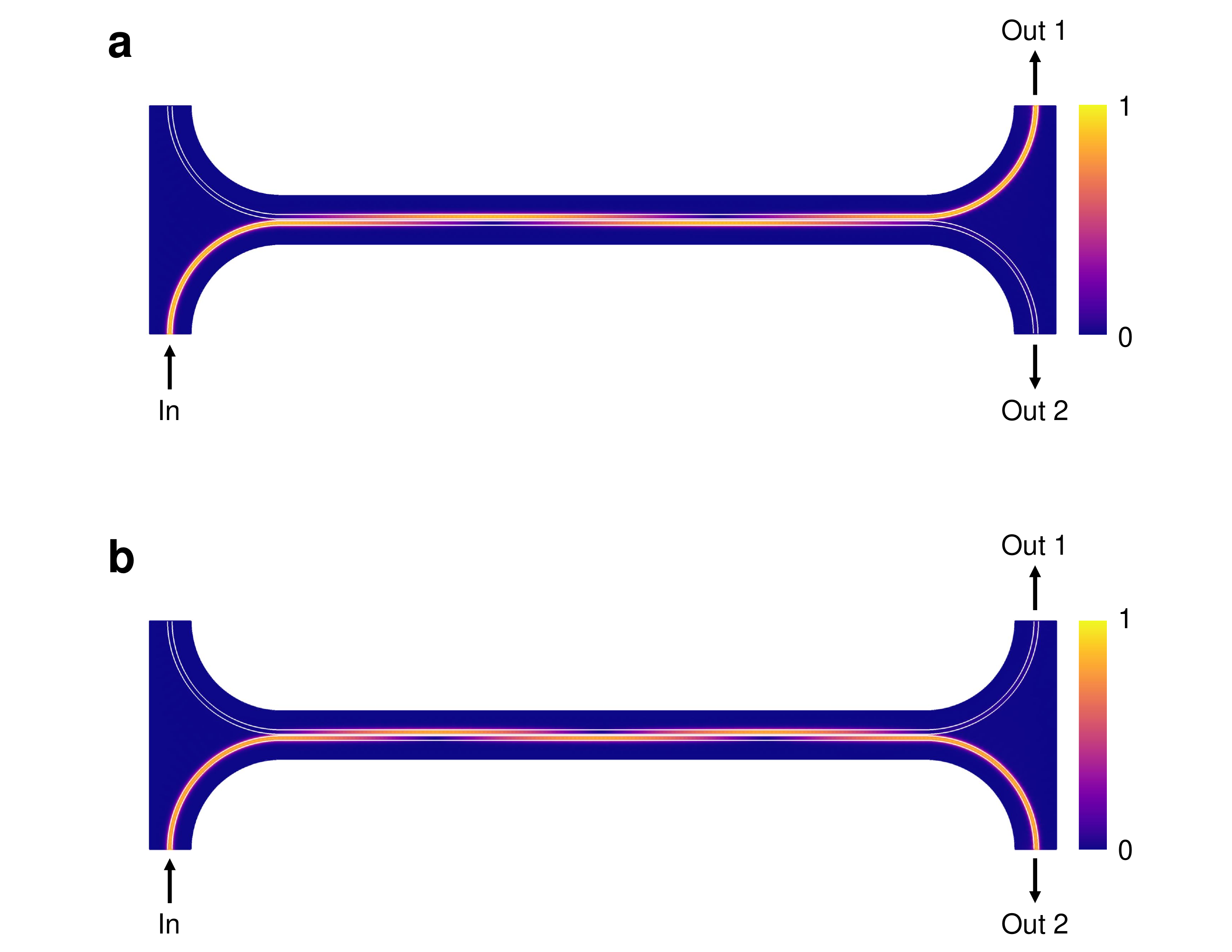}
\caption{\label{fig:Directional coupler}{Electric field plot of a non-actuated \textbf{(a)} and actuated \textbf{(b)} directional coupler.}}
\end{figure*}

The $2\times2$ add-drop network includes four switching elements each comprised of a directional coupler driven by two electrostatic comb-drive actuators \cite{Tsoukalas:20} (Fig.~\ref{fig:directional coupler fab}a). The directional couplers comprise two suspended waveguides in which the coupling between them and thus the flow of photons is controlled by an adjustable gap (Fig.~\ref{fig:directional coupler fab}b). The length and the initial gap of the directional coupler are chosen such that at zero bias voltage the input light is coupled to output 1 (Fig.~\ref{fig:Directional coupler}a). By driving the comb-drive actuators, the gap and thus the coupling strength is adjusted such that the input light is coupled to output 2 (Fig.~\ref{fig:Directional coupler}b). We use a revised version of the silicon bridge in the add-drop switch to provide higher optical transmission \cite{Papon:19}.

\begin{figure*}
\centering
\includegraphics[width=0.8\linewidth]{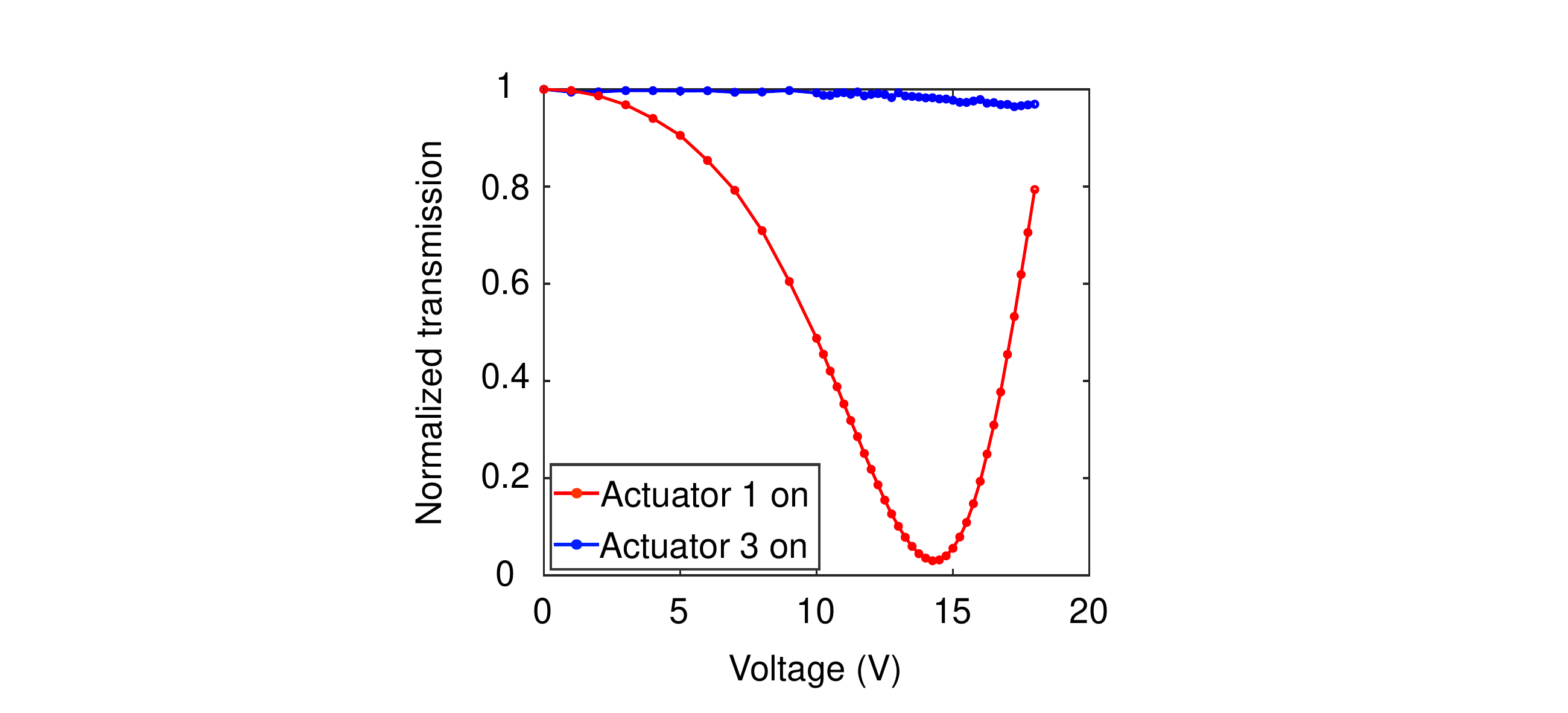}
\caption{\label{fig:Leakage}{\textbf{Characterization of cross-talk in nanoelectromechanical switch network based on EPCCs.} Measured transmission from $I_\text{2}$ to $T_\text{2}$ as a function of actuation voltage for two actuated comb drives showing no significant electrical crosstalk between the electrostatic actuators.}}
\end{figure*}

We characterize the electrical crosstalk between different nanoelectromechanical actuators by routing light through parts of the switch network while applying a voltage to another actuator outside the optical path and measuring any spurious influence on the transmission. In detail, the driving signal is applied first to the bottom-right actuator (labeled as 1 in Fig.~3d) and then to the top-left actuator (labeled as 3 in Fig.~3d), while the input light is coupled to $I_\text{2}$ and the output light is collected from $T_\text{2}$. Figure~\ref{fig:Leakage} shows the switching operation for this experiment. When actuator 1 is driven, the input light is routed from $T_\text{2}$ to $D_\text{2}$ (see Fig.~3h of the main text), so $T_\text{2}$ depends on the actuation voltage. When actuator 3 is driven, $T_\text{2}$ does not depend on the actuation voltage showing that there is no significant crosstalk from actuator 3 to actuators controlling $T_\text{2}$, i.e., actuators 1 and 4. Note that in our particular experimental setup, only one actuator (in this case actuator 3) is driven and the other actuators are at floating potentials, which explains the small deviation in $T_\text{2}$ at high voltages shown in Fig.~\ref{fig:Leakage}.

\FloatBarrier


\textcolor{black}{\bibliography{scibib,revtex-custom}}

\clearpage